\documentclass[12pt]{elsarticle}
\usepackage{natbib}
\usepackage{changes}
\usepackage{booktabs} 
\usepackage{graphicx} 
\usepackage{amsmath} 
\usepackage{enumerate}
\usepackage{epsfig}
\usepackage[export]{adjustbox}
\usepackage{amssymb}
\usepackage[utf8]{inputenc}
\usepackage{listings}
\usepackage{color, colortbl}
\usepackage{graphicx}
\usepackage{svg}
\usepackage{epstopdf}
\usepackage{caption}
\usepackage{amsmath,empheq,amssymb,mathtools}
\usepackage{gensymb}
\usepackage{float}
\usepackage{subfloat}
\usepackage{subfig}
\usepackage{mathtools}
\usepackage{cancel}
\usepackage{moreverb,url}
\usepackage{physics}
\usepackage{xcolor}	
\usepackage{hyperref}
\usepackage{tikz}
\usepackage{siunitx}
\usepackage{steinmetz}
\usepackage{multirow}
\usepackage{slashbox}
\usepackage[shortlabels]{enumitem}
\newtheorem{theorem}{Theorem}[section]
\newtheorem{corollary}{Corollary}[theorem]

\begin{document}
	
\begin{frontmatter}		

\title{Loop-shaping for reset control systems \\ \small{A higher-order sinusoidal-input describing functions approach} 
\footnote{Corresponding author: S. Hassan HosseinNia, Department of Precision and Microsystems Engineering, Delft University of Technology, Mekelweg 2, 2628 CD Delft, The Netherlands. Email: \url{s.h.hosseinniakani@tudelft.nl}}} 
		
\author[label1]{Niranjan Saikumar}
\author[label2]{Kars Heinen}
\author[label1]{and S. Hassan HosseinNia}
\address[label1]{Department of Precision and Micro System Engineering, Delft University of Technology, The Netherlands}
\address[label2]{Delft Center for Systems and Control, Delft University of Technology, The Netherlands}

\begin{abstract}
{
	The ever-growing demands on speed and precision from the precision motion industry have pushed control requirements to reach the limitations of linear control theory. Nonlinear controllers like reset provide a viable alternative since they can be easily integrated into the existing linear controller structure and designed using industry-preferred loop-shaping techniques. However, currently, loop-shaping is achieved using the describing function (DF) and performance analysed using linear control sensitivity functions not applicable for reset control systems, resulting in a significant deviation between expected and practical results. We overcome this major bottleneck to the wider adaptation of reset control with two contributions in this paper. First, we present the extension of frequency-domain tools for reset controllers in the form of higher-order sinusoidal-input describing functions (HOSIDFs) providing greater insight into their behaviour. Second, we propose a novel method which uses the DF and HOSIDFs of the open-loop reset control system for the estimation of the closed-loop sensitivity functions, establishing for the first time - the relation between open-loop and closed-loop behaviour of reset control systems in the frequency domain. The accuracy of the proposed solution is verified in both simulation and practice on a precision positioning stage and these results are further analysed to obtain insights into the tuning considerations for reset controllers.
}
\end{abstract}
		
		
\end{frontmatter}
	
\section{Introduction}

PID and the like linear controllers continue to dominate industrial control including the high-tech industry with precision applications such as photolithography wafer scanners, atomic force microscopes, adaptive optics etc. This status quo is likely to continue as observed in \cite{samad2019ifac}. An important reason for this sustained trend especially in the precision industry is that these linear controllers lend themselves for loop-shaping based design using the plant frequency response function (FRF) and for performance prediction using sensitivity functions in the frequency domain. However, the constant push for higher bandwidths, tracking precision, robustness cannot be met by linear controllers which are fundamentally limited by the waterbed effect \cite{bode1945network}. While nonlinear control theory has developed significantly over the decades, controllers compatible with well-established industry-standard techniques, especially design, prediction and analysis in the frequency domain, are required to meet future needs.

Reset control, first proposed by J. C. Clegg in 1958 \cite{clegg1958nonlinear}, is one such nonlinear control technique with significant potential to replace PID and its family of controllers. Reset technique was introduced for an integrator wherein its state is reset to zero when the error input hits zero. Describing function (DF) analysis of this element - reset integrator or more popularly dubbed as `Clegg Integrator (CI)' shows that CI has similar gain behaviour compared to a linear integrator, but with a significant phase advantage of only $38^\circ$ lag compared to $90^\circ$ in the linear case. 

This idea was extended in the form of `First order reset element (FORE)' in \cite{horowitz1975non,krishnan1974synthesis}, adding much needed tuning flexibility, with closed-loop performance improvement using reset control also shown for the first time in the same works. Over the years, elements such as `Second-order reset element (SORE)' \cite{hazeleger2016second} and `Fractional-order reset element (FrORE)' \cite{saikumargeneralized} have been introduced expanding the design freedom. Additional degrees of tuning have also been introduced with the PI+CI \cite{banos2007definition} and partial reset techniques \cite{beker2004fundamental}, with the latter resulting in generalized reset elements \cite{saikumar2019constant}. The advantage of reset control in improving performance has been extensively studied from process to motion control systems \cite{banos2011reset,chen2001analysis,zheng2000experimental,hosseinnia2013fractional,beker2001plant,wu2007reset,palanikumar2018no,chen2018beyond,chen2019development,akyuz2019reset}. While the mentioned works have retained the original condition of resetting the state when the error input hits zero, several works have also looked at modifying the reset condition to gain performance improvements \cite{li2011optimal,panni2012set}. However, most of these alternative reset conditions do not lend themselves for frequency domain analysis and hence are not the focus of this study.

While a large volume of work exists on the use of traditional reset technique in practice, a large fraction of this has been limited to the exploitation of the reduced phase lag advantage and hence reset is mainly used in the integrator part of PID. Recently, we introduced the `Constant-in-gain Lead-in-phase (CgLp)' \cite{saikumar2019constant,saikumar2019resetting} element aimed at a more wholistic utilization of reset from a loop-shaping perspective to gain significant improvements in tracking precision, bandwidth and stability. This CgLp element can potentially replace the derivative part of PID and go beyond \cite{valerio2019reset,saikumar2019complex}. While the potential of reset control to go beyond the limitations of linear control has been well-established, a fundamental roadblock which remains is the lack of a clear frequency domain analysis method for reset control systems, which is critical for design, performance prediction and analysis in the loop-shaping framework. The current use of DF for loop-shaping design falls short especially in precision systems and we have reported a large deviation from performance estimated using linear analysis of DF in our previous works \cite{saikumar2019constant,akyuz2019reset,saikumar2019complex}.

In this paper, we attempt to clear this bottleneck through two contributions for analysing performance in the frequency-domain. First, we provide the extension of a frequency-domain tool called `Higher-order sinusoidal-input describing functions (HOSIDFs)' for reset controllers enabling us to do a deeper analysis in the open-loop. Second, we propose a method which allows us to translate the open-loop behaviour to closed-loop in the frequency domain, which in essence are the sensitivity functions for reset control systems. The remainder of this paper is structured as follows. The preliminaries of reset control along with existing describing function analysis method are presented in Section \ref{section_Reset_control}. The HOSIDF tool as applied to reset controllers is presented in Section \ref{section_Hosidf}, followed by the novel method establishing sensitivity functions for reset control systems in Section \ref{section_Sensitivity}. The accuracy of the proposed solutions is tested in both simulation and practice on a precision motion system in Section \ref{section_validation}, followed by a general analysis of the results and a discussion on their implication for loop-shaping of reset controllers in Section \ref{section_Analysis}. The paper is concluded with a summary and remarks for future work in Section \ref{section_Conclusions}.

\section{Preliminaries on Reset control}
\label{section_Reset_control}

The preliminaries of reset control including definition, describing function, reset elements, stability and the problem with using DF for loop-shaping are presented in this section.

\subsection{Definition of Reset controller}
\label{subsection_Reset_controller}

While reset controllers with various state/input/time dependent resetting conditions/laws exist in literature, the most popular reset law which lends itself for frequency domain analysis is based on the input (generally error) hitting zero. This is also referred to as `zero-crossing law'. A SISO reset controller with this law can be defined using the following equations:
\begin{align} 
	\label{eq_reset_controller}
	\mathcal{R} =
	\begin{cases}
		{\dot{x}}_R(t) = A_R x_R(t) + B_R e(t) & \ e(t)\neq 0 \\
		x_R(t^+) = A_\rho x_R(t)& \ e(t)=0 \\
		u_R(t) = C_R x(t) + D_R e(t) \\
	\end{cases}
\end{align}
where $e(t)$ is the error input, $u_R(t)$ is the controller output and $x_R(t) \in \mathbb{R}^{n_\mathcal{R}}$. $A_R$, $B_R$, $C_R$, and $D_R$ represent the state-space matrices and are together referred to as the base-linear controller. The first equation provides the non-reset continuous dynamics referred to as flow dynamics, whereas the resetting action is given by the second equation referred to as the jump dynamic. $A_\rho$ is the resetting matrix which determines the after-reset values of the states and is generally of form $diag(\gamma_1,\gamma_2,....,\gamma_{n_\mathcal{R}})$ where $\gamma_i \in [-1,1]$. A general reset controller can be defined using (\ref{eq_reset_controller}) to include the linear non-resetting controller part in which case, the first $n_r$ states are the resetting states, followed by $n_{nr}$ non-resetting states, with $n_\mathcal{R} = n_r + n_{nr}$. In this case, the resetting matrix $A_\rho$ can be represented as
$$
A_\rho = \begin{bmatrix} A_{\rho_r} & \\ & I \end{bmatrix}
$$

\subsection{Describing function (DF)}
\label{subsection_DF}

Reset controllers $\mathcal{R}$ are analysed in the frequency domain through the sinusoidal input describing function (DF), which considers only the first harmonic of the Fourier series expansion of the periodic output $u_R(t)$ to a sinusoidal input $e(t)$. The analytical equations for the calculation of DF assuming $A_\rho$ is Schur stable (which establishes convergence in open-loop) are provided in \cite{guo2009frequency} as
\begin{equation}
	\label{eq_DF}
	H_1(\omega) = C_R (j\omega I-A_R)^{-1} (I+j\Theta_D (\omega))  B_R+D_R
\end{equation}
where
\begin{eqnarray*}
	\left.\begin{aligned}
		&\Lambda(\omega) = \omega^2 I+A_R^2 \\
		&\Delta(\omega) = I+e^{\big(\tfrac{\pi}{\omega} A_R\big)} \\
		&\Delta_r (\omega) = I+A_\rho e^{\big(\tfrac{\pi}{\omega} A_R\big)} \\
		&\Gamma_r (\omega) = \Delta_r^{-1} (\omega)  A_\rho  \Delta(\omega)  \Lambda^{-1} (\omega) \\
		&\Theta_D (\omega) = \dfrac{-2\omega^2}{\pi}  \Delta(\omega)  \left[ \Gamma_r (\omega)- \Lambda^{-1} (\omega)\right]
	\end{aligned}\right.
\end{eqnarray*}

\subsection{Reset elements}
\label{subsection_reset_elements}

The most popular and relevant reset elements are presented here.

\subsubsection{Generalized Clegg Integrator (GCI)} 

The first reset element as introduced by Clegg in \cite{clegg1958nonlinear} can be generalized with partial reset allowing for the integrator state to be reset to a fraction of its value instead of zero. This is represented in transfer function form as below with the arrow indicating reset.
\begin{align} 
	\label{eq_gci}
	\text{GCI}= \dfrac{1}{\cancelto{A_\rho}{\alpha s}}
\end{align}
where $\alpha$ corrects for the change in gain of DF seen at all frequencies. This is noted to be $1.62$ for $\gamma = 0$ in literature and varies for different values of $\gamma$. $A_\rho = \gamma \in [-1,1]$ allows for the generalization of Clegg Integrator. The corresponding state-space matrices as per (\ref{eq_reset_controller}) are given as
\begin{equation}
	A_R = 0,\quad B_R = 1/\alpha,\quad C_R = 1,\quad D_R = 0 \nonumber
\end{equation}

\subsubsection{Generalized FORE (GFORE)} 

FORE presented in \cite{horowitz1975non} was generalized and extended as GFORE by \cite{guo2009frequency} creating a first-order reset filter with the resetting matrix $A_\rho$ controlling the level of reset. GFORE with corner frequency at $\omega_r$ can be represented as
\begin{align} 
	\label{eq_gfore}
	\text{GFORE}= \dfrac{1}{\cancelto{A_\rho}{\frac{s}{\alpha \omega_r} +1}}
\end{align}
where $\alpha$ accounts for the change in the gain of GFORE at high frequencies as noted in \cite{saikumar2019constant}, $A_\rho = \gamma \in [-1,1]$ with the value of $\alpha$ dependent on the value of $\gamma$. The corresponding state-space matrices as per (\ref{eq_reset_controller}) are given as
\begin{equation}
	A_R = -\alpha\omega_r,\quad B_R = \alpha\omega_r,\quad C_R = 1,\quad D_R = 0 \nonumber
\end{equation}

\subsubsection{Generalized SORE (GSORE)} 

SORE creates a second-order reset filter and allows for additional tuning of the damping parameter of the filter. SORE presented in \cite{hazeleger2016second} was generalized in \cite{saikumar2019constant} and can be represented as:
\begin{equation} 
	\label{eq_gsore}
	\text{GSORE}= 
	\dfrac{1}{\Big(\cancelto{A_\rho}{\frac{s}{\alpha \omega_r}\Big)^2 + 2\kappa \beta_r \frac{s}{\alpha \omega_r}+1}}
\end{equation}
where $\alpha$ again corrects for the change in gain, $\beta_r$ being the damping coefficient, $\kappa$ being the correction factor for the change in damping coefficient and resetting matrix $A_\rho = \gamma I$ with $\gamma \in [-1,1]$. The corresponding state-space matrices as per (\ref{eq_reset_controller}) are given as
\begin{align*}
	A_R =
	\begin{bmatrix} 0 & 1 \\ -(\alpha \omega_r)^2 & - 2\kappa \alpha \beta_r \omega_r \end{bmatrix}, \
	& B_R =
	\begin{bmatrix} 0   \\  (\alpha \omega_r)^2 \end{bmatrix}, \\
	C_R =
	\begin{bmatrix} 1 &  0 \end{bmatrix}, \
	& D_R = 0
\end{align*}

\subsection{Stability of reset control systems}
\label{subsection_stability}

\begin{figure} 
	\centering
	\includegraphics[trim = {0 0 0 0}, width=1\linewidth]{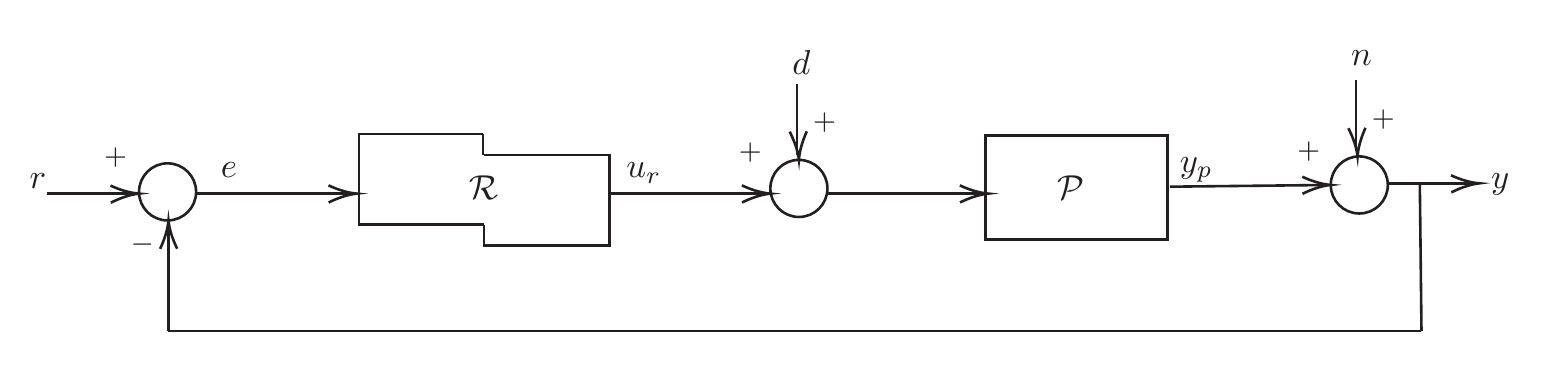}
	\caption{Reset Control System with linear plant $\mathcal{P}$ and reset feedback controller $\mathcal{R}$ with reference $r$, process noise $d$ and measurement noise $n$.}
	\label{fig_RCS}
\end{figure}

Consider $\mathcal{R}$ in closed-loop with a linear plant $\mathcal{P}$ as shown in Fig. \ref{fig_RCS} having state-space matrices $A_p$, $B_p$, $C_p$ and $x_p \in \mathbb{R}^{n_\mathcal{P}}$ such that
\begin{align} 
	\label{eq_plant}
	\mathcal{P} =
	\begin{cases}
		{\dot{x}}_p(t) = A_p x_p(t) + B_p u_R(t) \\
		y_p(t) = C_p x_p(t) \\
	\end{cases}
\end{align}
Neglecting exogenous signals, $r$, $d$ and $n$, combining (\ref{eq_reset_controller}) and (\ref{eq_plant}) gives
\begin{align} 
	\label{eq_RCS}
	\mathcal{RCS} =
	\begin{cases}
		{\dot{x}}(t) = A x(t) & x \notin \mathcal{J}\\
		x^+ = Rx & x \in \mathcal{J}\\
		y(t) = C x(t)
	\end{cases}
\end{align}
where $x^T = \begin{bmatrix}{x^T_R} & {x^T_p}\end{bmatrix}^T \in \mathbb{R}^{n}$ with $n = n_\mathcal{R} + n_\mathcal{P}$,
$$
A =
\begin{bmatrix} A_R & -B_RC_p \\ B_pC_R & A_p \end{bmatrix}, \
C =
\begin{bmatrix} 0 & C_p \end{bmatrix}
$$
$$
R = \begin{bmatrix} A_\rho & \\ & I \end{bmatrix} \text{and  } \mathcal{J} := \{x \in \mathbb{R}^{n} | Cx = 0\}
$$

The stability of this closed loop reset control system ($RCS$) can be verified using the $H_\beta$ condition provided in \cite{beker2004fundamental}.
\begin{theorem}
	\label{th_stability}
	The $RCS$ (\ref{eq_RCS}) is quadratically stable if and only if the $H_\beta$ condition holds, i.e., there exists a $\beta \in \mathbb{R}^{n_r}$ and a positive definite matrix $P_r \in \mathbb{R}^{n_r\times n_r}$ such that the transfer function 
	\begin{equation}
		\label{eq_H_beta}
		H_\beta(s) := \begin{bmatrix} P_r & 0_{n_r\times n_{nr}} & \beta C_p \end{bmatrix}(sI - A)^{-1}\begin{bmatrix}I_{n_r} \\ 0\end{bmatrix}
	\end{equation}
	is strictly positive real and additionally a non-zero reset matrix $A_{\rho_r}$ satisfies the condition
	\begin{equation}
		\label{eq_Arho}
		A^T_{\rho_r}P_r A_{\rho_r} - P_r \leq 0
	\end{equation}
\end{theorem}

BIBO stability of $\mathcal{RCS}$ if also guaranteed if the above conditions are met \cite{beker2004fundamental}.

\subsection{CgLp-PID design and the problem of describing function}
\label{subsection_CgLpPID_DF}

We introduced the `Constant-in-gain Lead-in-phase' (CgLp) element in \cite{saikumar2019constant} to provide broadband phase compensation. This is done by combining a GFORE or GSORE element with corner frequency $\omega_r$ in series with a corresponding first or second order linear lead element with zero and pole located at $\omega_r$ and $\omega_f$ ($\omega_f >> \omega_r$) respectively. While the resetting action results in minor changes to the gain profile of the reset element (compensated by $\alpha$ as noted in \ref{subsection_reset_elements}), it results in a substantial reduction of phase lag as seen in the DF. This gain profile is cancelled by that of the linear lead element to provide constant gain while the linear phase lead obtained combined with the reduced phase lag of the reset element results in broadband phase lead in the range $[\omega_r,\omega_f]$. In reality, phase compensation can be obtained even at frequencies below $\omega_r$. The state-space matrices of CgLp created using a GFORE along with a first order linear lead are given as
\begin{align*}
	A_{CgLp} =
	\begin{bmatrix} -\alpha\omega_r & 0 \\ \omega_f & -\omega_f \end{bmatrix}, \
	& B_{CgLp} =
	\begin{bmatrix} \alpha\omega_r   \\  0 \end{bmatrix}, \\
	C_{CgLp} =
	\begin{bmatrix} \frac{\omega_f}{\omega_r} &  \Big(1 - \frac{\omega_f}{\omega_r}\Big) \end{bmatrix}, \
	& D_{CgLp} = 0,
\end{align*}
\begin{equation*}
	A_{\rho_r} = \begin{bmatrix} \gamma & 0\\0 & 1\end{bmatrix}
\end{equation*}

The design of the CgLp-PID is done in two main steps. In the first step, the linear PID controller is designed using loop-shaping with the frequency response function (FRF) of $\mathcal{P}$ to meet the performance specifications in terms of tracking, steady-state precision, disturbance rejection. While the closed-loop system has to be stable, the phase margin (PM) requirement related to stability is ignored in this first step. In the second step, a CgLp element is designed to provide phase compensation and obtain the required PM as per DF. The series combination of CgLp with PID results in CgLp-PID controller design. More details on this can be found in \cite{saikumar2019constant,saikumar2019resetting}.

\begin{figure}
	\centering
	\includegraphics[trim = {1.0cm 0.5cm 1.5cm 1cm}, width=1\linewidth]{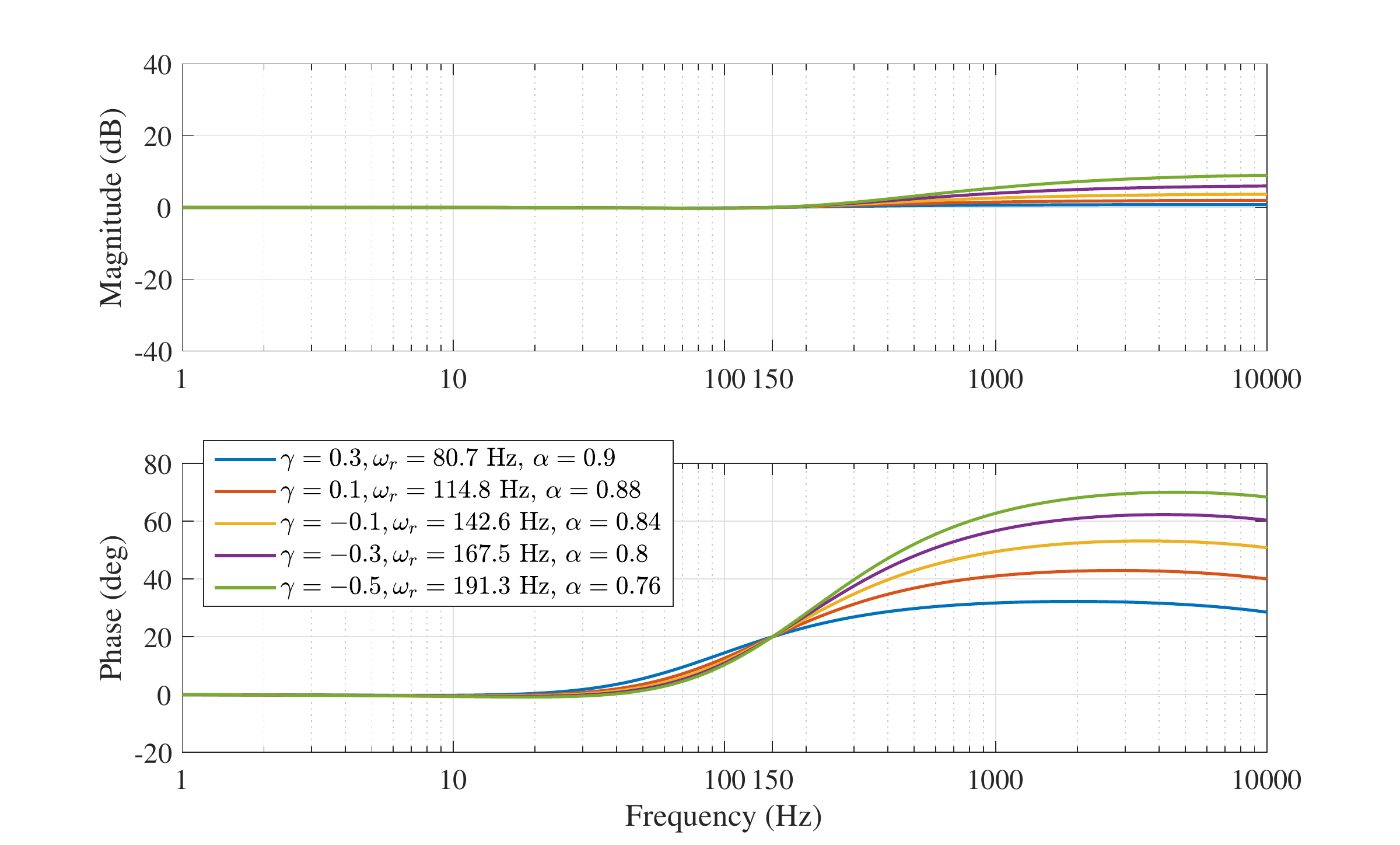}
	\caption{Describing function of multiple CgLp elements designed to provide $20^\circ$ phase compensation at $150\ Hz$. Slight deviation from unity gain is seen due to the nonlinear frequency behaviour of the GFORE.}
	\label{fig_DF_CgLps}	
\end{figure}

The phase compensation of CgLp is seen through the DF analysis and assuming $\omega_f >> \omega_r$, the two variables $\omega_r$ and $\gamma$ are the tuning knobs of this element. Since CgLp is capable of providing large phase compensation of up-to $52^\circ$ with a traditional $\gamma = 0$, phase compensation in general can be achieved with several different combinations of $\{\omega_r, \gamma\}$ as shown in Fig. \ref{fig_DF_CgLps} for $20^\circ$ phase compensation at $150\ Hz$. 

Now, consider the plant $\mathcal{P}$ given by
\begin{equation}
	\label{eq_plant_spider}
	\mathcal{P} = \dfrac{6.615e5}{83.57s^2 + 279.4s + 5.837e5}
\end{equation}
A PID controller is designed as given below to obtain a gain cross-over frequency of $150\ Hz$ with a phase margin of $20^\circ$.
\begin{equation}
	\label{eq_PID_example}
	\text{PID}(s) = \text{K}\Bigg(1 + \dfrac{\omega_i}{s}\Bigg)\Bigg(\dfrac{\frac{s}{\omega_d} + 1}{\frac{s}{\omega_t} + 1}\Bigg)\Bigg(\dfrac{1}{\frac{s}{\omega_{lpf}} + 1}\Bigg)
\end{equation}
where $\omega_i = 2\pi15, \omega_{lpf} = 2\pi1500, \omega_d = 2\pi84.34, \omega_t = 2\pi266.75\ rad/s, \text{K} = 60.835$

\begin{figure}
	\centering
	\includegraphics[trim = {1.0cm 0.5cm 1.5cm 1cm}, width=1\linewidth]{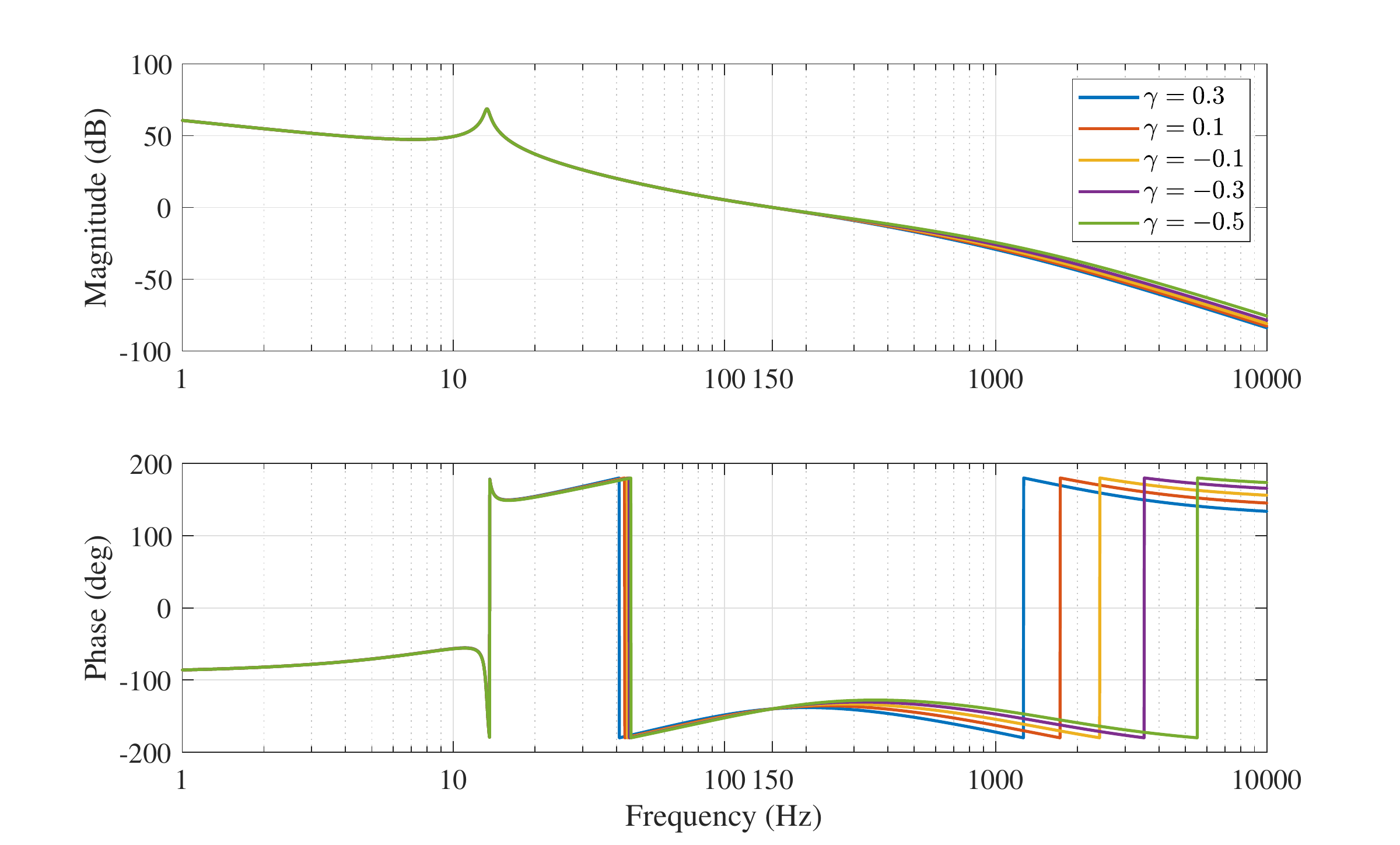}
	\caption{Describing function of open-loop with the 5 CgLp compensators of Fig. \ref{fig_DF_CgLps} used to design 5 different CgLp-PIDs with same PM of $40^\circ$.}
	\label{fig_DF_OLs}	
\end{figure}

\begin{figure}
	\centering
	\includegraphics[trim = {1.0cm 0.5cm 1.5cm 1cm}, width=1\linewidth]{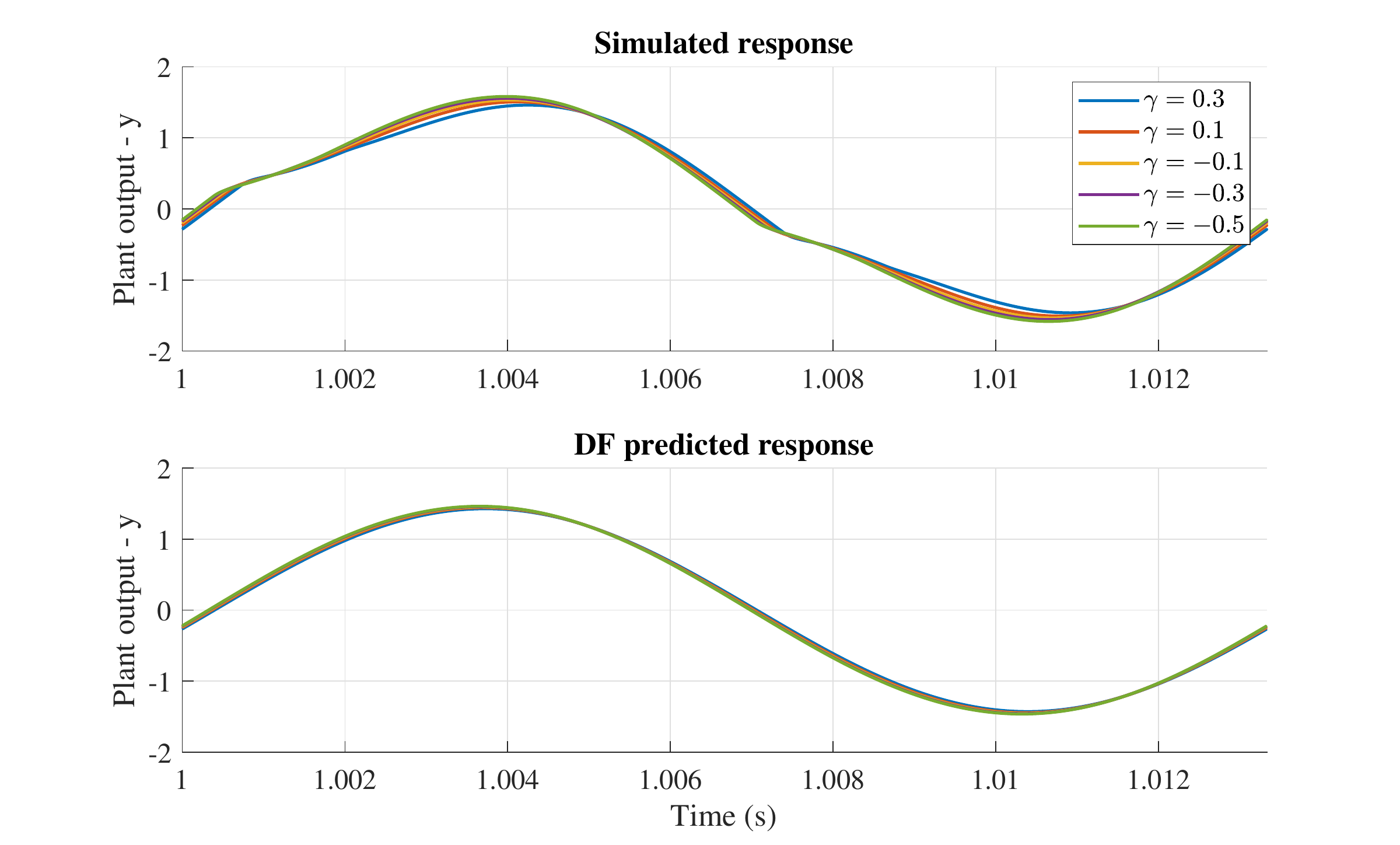}
	\caption{Steady state responses to a sinusoidal reference of $\SI{75}{Hz}$ for the closed-loop systems whose open-loop DFs are shown in Fig. \ref{fig_DF_OLs}.}
	\label{fig_step_response_problem}	
\end{figure}

The various CgLp compensators of Fig. \ref{fig_DF_CgLps} are used to make 5 different CgLp-PID controllers such that the DF of the open-loop now shows a PM of $40^\circ$ as shown in Fig. \ref{fig_DF_OLs} with the steady state responses to a sinusoidal excitation as reference for all 5 systems shown in Fig. \ref{fig_step_response_problem}. The responses as predicted by DF are also shown. The simulated responses clearly show that the plant output is not a single sinusoid, and additionally the difference in peak output between the 5 systems is not captured by the DF predicted output. Similar differences in performance between different CgLp-PID controllers and additionally deviation from DF based predicted performance in tracking and precision are noted in greater detail in \cite{saikumar2019constant}. Additionally, the presence of limit cycles resulting in large errors (not predicted by DF) when the integrator is reset is well recorded in literature \cite{banos2011reset}. This clearly establishes the problem associated with the exclusive use of DF for design and analysis of $\mathcal{RCS}$ and the requirement of more tools for the frequency domain analysis of these systems.

\section{Higher-order Sinusoidal-input Describing function (HOSIDF) for Reset controllers}
\label{section_Hosidf}   

Frequency domain-based concepts and tools like loop-shaping which use the FRF of the plant assume linear system behaviour. Although in this paper we deal with linear systems, the use of nonlinear reset control for performance improvement is handled in literature through the quasi-linear descriptor of the describing function. However, the exclusive use of DF is highly dependent on the first component of the Fourier series expansion dominating the other components, which is not true for a large class of reset controllers. Additionally, vital information regarding the system behaviour is neglected. \cite{nuij2006higher} introduces the concept of a virtual harmonic generator as a bridge between the frequency domain analysis of linear and a class of static nonlinear dynamic systems to extend DF for higher-order functions resulting in higher-order sinusoidal input describing functions (HOSIDFs). In this section, we apply these concepts to reset controllers for open-loop frequency domain analysis and present the analytical equations for the calculation of these functions.

\subsection{Virtual harmonic generator}
\label{subsection_virtual_harmonic_gen}

\begin{figure*} 
	\centering
	\includegraphics[trim = {0 0 0 0}, width=1\linewidth]{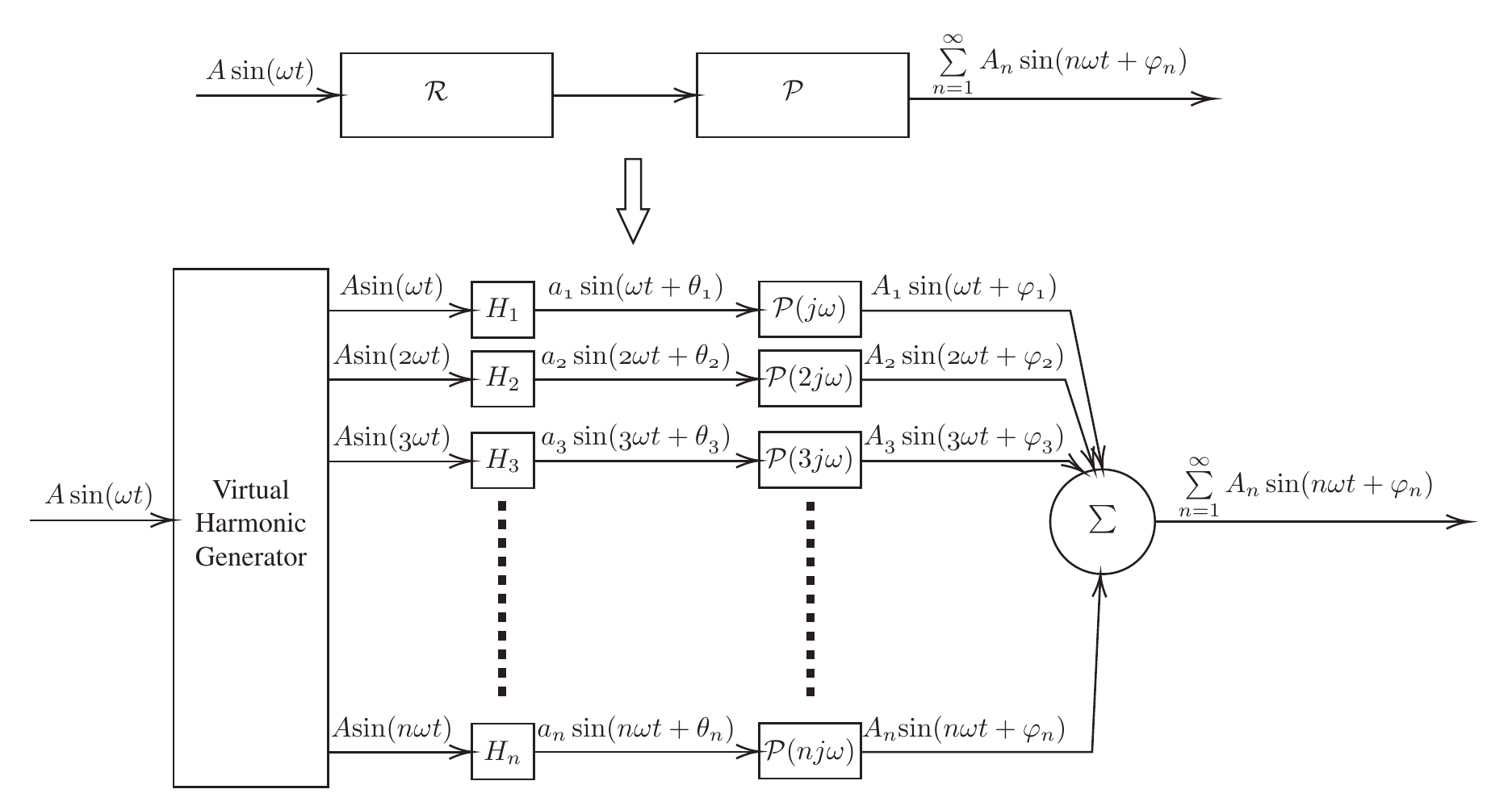}
	\caption{Representation of Higher-order sinusoidal-input describing function for open-loop reset control systems} 
	\label{fig_HOSIDF_principle}
\end{figure*}

Reset controllers $\mathcal{R}$ are nonlinear time-invariant systems and their stability and convergence in open-loop is ensured if $A$ is Hurwitz and $A_\rho$ is Schur stable \cite{guo2009frequency}. For $e(t) = A\sin(\omega t)$ input signal, the steady-state output $u_R(t)$ is periodic and consists of harmonics of the fundamental frequency $\omega$ and hence can be expressed as the summation of harmonics of the input signal, with exclusive amplitude and phase associated with each harmonic. Since reset is not an amplitude-dependent nonlinearity, this system can be modelled as a virtual harmonics generator and a linear system associated with each harmonic according to \cite{nuij2006higher}, where the generator converts the input signal into a harmonic signal consisting of an infinite number of harmonics. The plant $\mathcal{P}$ can also be included in this model as shown in Fig. \ref{fig_HOSIDF_principle} as a complete open-loop model of $\mathcal{RCS}$. As seen, since a separate amplitude and phase is associated with each harmonic, an exclusive linear block is modelled for each harmonic with a parallel interconnection. Since we are considering linear plants, this essentially results in a modification of the Hammerstein model \cite{narendra1966iterative}.

\subsection{DF and HOSIDF}
\label{subsection_DF_HOSIDF}

The describing function $H_1(\omega)$ of a system is defined as the ratio of the fundamental component of $u_R(t)$ and the input $e(t)$. This describing function can be considered as the first element of a set of higher-order describing functions $H_n(\omega)$, where each function is the complex ratio of the $n^{th}$ harmonic of the output $u_R(t)$ to the input $e(t)$. Hence as per Fig. \ref{fig_HOSIDF_principle}, higher-order describing function of $\mathcal{R}$ can be calculated as
\begin{equation}
	\label{eq_Fourier}
	H_n(\omega) = \frac{a_n(\omega)e^{j(n\omega t + \theta_n(\omega))}}{A}
\end{equation}
Note that in this case, $\omega$ refers to the fundamental frequency of the output, i.e., the frequency of the input signal, while the frequency of the harmonic is $n\omega$.

The describing function $H_1(\omega)$ of a reset controller can be analytically calculated as per the equations provided in \cite{guo2009frequency} and repeated in \ref{subsection_DF}. Based on this work, the equations to analytically calculate the HOSDIF of a reset controller $\mathcal{R}$ are provided next.

\begin{theorem}
	\label{th_HOSIDFs}
	For a reset controller $\mathcal{R}$,
	\begin{equation}
		\label{eq_HOSIDF}
		H_n(\omega) = \begin{cases}
			C_R (j\omega I-A_R)^{-1} (I+j\Theta_D (\omega))  B_R+D_R \\
			\hspace{4.5cm} \text{for } n = 1\\
			C_R (jn\omega I-A_R)^{-1} (I+j\Theta_D (\omega))  B_R \\
			\hspace{4.5cm} \text{for odd } n \geq 2 \\
			0 \\
			\hspace{4.5cm} \text{for even } n \geq 2
		\end{cases}
	\end{equation}
	with 
	\begin{eqnarray}
		\label{eq_HOSIDF_part_eq}
		\left.\begin{aligned}
			&\Lambda(\omega) = \omega^2 I+A_R^2 \\
			&\Delta(\omega) = I+e^{(\tfrac{\pi}{\omega} A_R)} \\
			&\Delta_r (\omega) = I+A_\rho e^{(\tfrac{\pi}{\omega} A_R)} \\
			&\Gamma_r (\omega) = \Delta_r^{-1} (\omega)  A_\rho  \Delta(\omega)  \Lambda^{-1} (\omega) \\
			&\Theta_D (\omega) = \dfrac{-2\omega^2}{\pi}  \Delta(\omega)  \left[ \Gamma_r (\omega)- \Lambda^{-1} (\omega)\right]
		\end{aligned}\right.
	\end{eqnarray}
\end{theorem}
\textit{Proof:} $\mathcal{R}$ is divided into the linear part consisting of the $D_R$ matrix and the nonlinear part consisting of the rest. We first analyse the nonlinear part of $\mathcal{R}$. (\ref{eq_HOSIDF_part_eq}) are defined for convenience. For a sinusoidal input $e(t) = \sin(\omega t)$ (amplitude normalised since reset is not an amplitude dependent nonlinearity), the steady-state output (for $D_R = 0$) can be calculated as given in \cite{guo2009frequency} as
\begin{align}
	\label{eq_steady_State}
	u_{ss}(t) = & C_Re^{A_Rt}\theta_k(\omega) \nonumber \\ & - C_R\Lambda^{-1}(\omega)[\omega I\cos(\omega t) + A_R\sin(\omega t))]B_R
\end{align}
where $\theta_k(\omega) = (-1)^{k+1}e^{-A_Rt_k}[\Gamma_r(\omega) - \Lambda^{-1}(\omega)]\omega B_R$ and $t \in (t_k, t_{k+1}]$ with $t_k = k\pi/\omega$ and $k = 0,1,2,\cdot \cdot \cdot \cdot \cdot$ providing the reset instants.

The Fourier series component for the first harmonic needed for the calculation of DF is provided in \cite{guo2009frequency} as noted in \ref{section_Reset_control}. Hence only higher orders are calculated here. The $n^{th}$ harmonic component of $u_{ss}(t)$ is given as
\begin{align*}
	U_{ss_n}(\omega) & =\frac{\omega}{2\pi} \int_{0}^{\frac{2\pi}{\omega}} u_{ss}(t)e^{-j\omega nt} dt\\
	& = \frac{\omega C_R}{2\pi} (I_1 + I_2) - \frac{\omega C_R \Lambda^{-1}(\omega)}{2\pi}(\omega J_1 + A_RJ_2)B_R
\end{align*}
where
\begin{align*}
	I_1 & = \int_{0}^{\frac{\pi}{\omega}}e^{A_Rt}\theta_0(\omega)e^{-j\omega nt} dt\\
	& = \theta_0(\omega)(A_R - j\omega nI)^{-1} (e^{\frac{\pi}{\omega}A_R}(-1)^n - 1)\\
	& = [\Gamma_r(\omega) - \Lambda^{-1}(\omega)]\omega B_R(A_R - j\omega nI)^{-1}(1 - e^{\frac{\pi}{\omega}A_R}(-1)^n)
\end{align*}
\begin{align*}
	I_2 & = \int_{\frac{\pi}{\omega}}^{\frac{2\pi}{\omega}}e^{A_Rt}\theta_1(\omega)e^{-j\omega nt} dt\\
	& = \theta_1(\omega)(A_R - j\omega nI)^{-1} (e^{\frac{2\pi}{\omega}A_R} - e^{\frac{\pi}{\omega}A_R}e^{-j\pi n})\\
	& = [\Gamma_r(\omega) - \Lambda^{-1}(\omega)]\omega B_R(A_R - j\omega nI)^{-1}(e^{\frac{\pi}{\omega}A_R} - e^{-j\pi n})
\end{align*}
\begin{align*}
	J_1 & = \int_{0}^{\frac{\pi}{\omega}}e^{-j\omega nt}\cos(\omega t) dt\\
	& = 0 \text{ for } n \geq 2
\end{align*}
\begin{align*}
	J_2 & = \int_{0}^{\frac{\pi}{\omega}}e^{-j\omega nt}\sin(\omega t) dt\\
	& = 0 \text{ for } n \geq 2
\end{align*}
Hence we get
\begin{align*}
	U_{ss_n} & = \frac{\omega C_R}{2\pi}(I_1 + I_2) \text{for } n \geq 2\\
	& = \frac{\omega C_R}{2\pi}[\Gamma_r(\omega) - \Lambda^{-1}(\omega)]\omega B_R(A-R - j\omega nI)^{-1}\\& \hspace{2cm} \times [1 - e^{\frac{\pi}{\omega}A_R}(-1)^{-1} + e^{\frac{\pi}{\omega}A_R} - e^{-j\pi n}]
\end{align*}
The last term of the above equation is 0 for even values of $n$ indicating that the steady-state output $\mathcal{R}$ is an odd function of time. Rewriting this, we get
\begin{align*}
	U_{ss_n} = \begin{cases*}
		& $\frac{\omega^2C_R}{\pi}(A_R - j\omega nI)^{-1}\Delta(\omega)[\Gamma_r(\omega) - \Lambda^{-1}(\omega)]B_R$\\
		& \hspace{4.5cm} \text{for odd} $n \geq 2$\\
		& 0\\
		& \hspace{4.5cm} \text{for even} $n \geq 2$
	\end{cases*}
\end{align*}
The linear part of the reset controller comprising purely of the $D_R$ matrix does not affect the harmonics ($n \geq 2$). However, it does affect the first harmonic. Combing these parts, the complete HOSIDF equations can be written as in (\ref{eq_HOSIDF}). Hence, proved.

From Fig. \ref{fig_HOSIDF_principle}, it can be seen that the parallel interconnection used to model $\mathcal{R}$ is also extended to include $\mathcal{P}$. However, although $\mathcal{P}$ is linear, the branch associated with the $H_n(\omega)$ of $\mathcal{R}$ has a sinusoidal input of frequency $n\omega$. Hence, the frequency response at $n\omega$ should be used.

\begin{corollary}
	\label{cor_ol_hosidf}
	For the reset controller $\mathcal{R}$ and linear plant $\mathcal{P}$, the open-loop HOSIDF is obtained as
	\begin{equation}
		\label{eq_openloop_hosidf}
		L_n(\omega) = \begin{cases}
			H_n(\omega)\mathcal{P}(n\omega) & \text{for odd } n\\
			0 & \text{for even } n
		\end{cases}
	\end{equation}
\end{corollary}

\subsection{Visualization of HOSIDF}
\label{subsection_Visualization_HOSIDF}

\begin{figure}
	\centering
	\includegraphics[trim = {1.0cm 0.5cm 1.5cm 1cm}, width=1\linewidth]{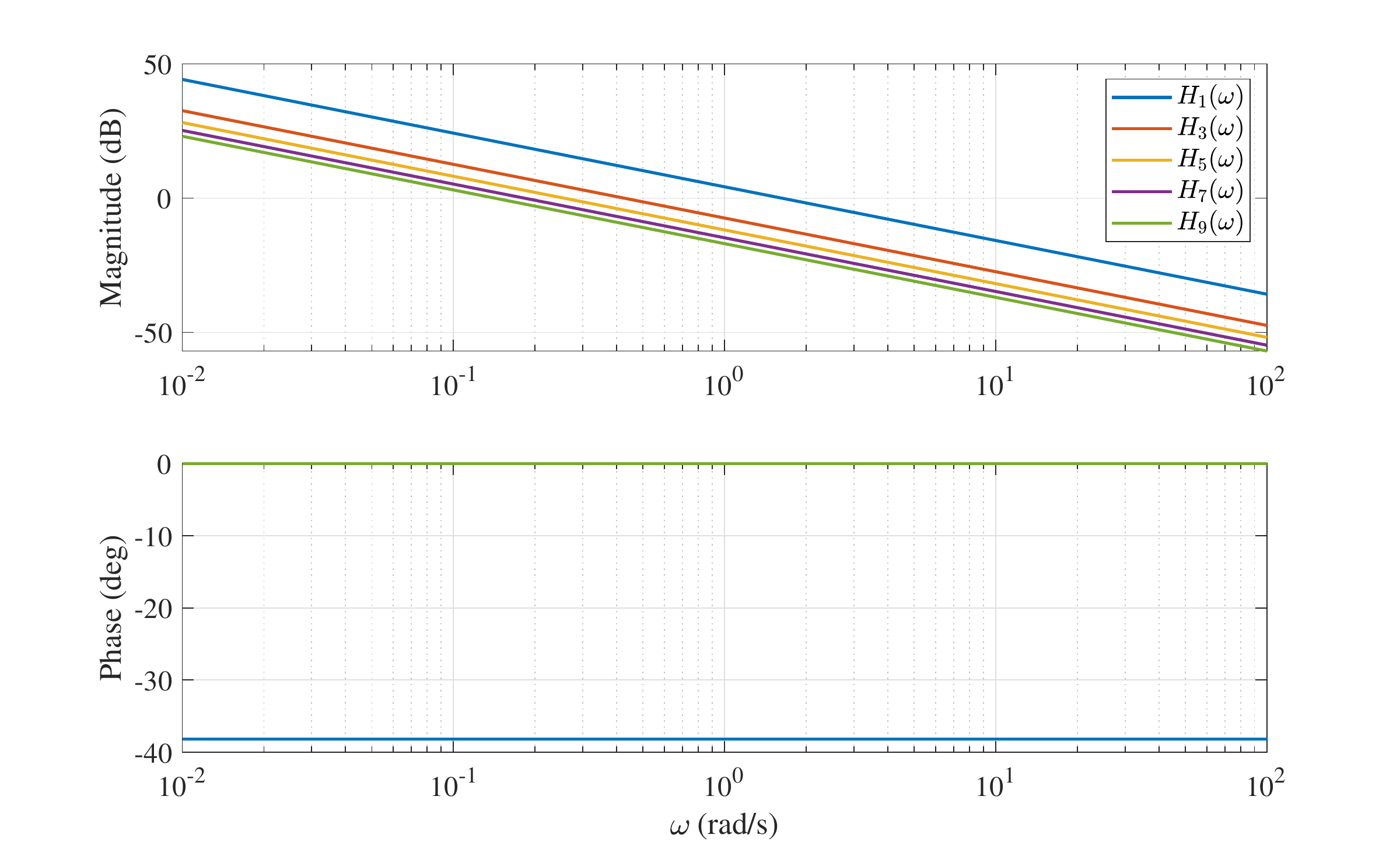}
	\caption{HOSIDFs of a Clegg integrator. $\angle H_{n>1}(\omega) = 0^\circ\ \ \forall \omega \in \mathbb{R}$.}
	\label{fig_HOSIDFs_Clegg}	
\end{figure}

The development of HOSIDF for $\mathcal{R}$ and the analytical equations (\ref{eq_HOSIDF}) allow for quick calculation and accurate representation of the frequency domain behaviour. The HOSIDF for a Clegg integrator are obtained and plotted in Fig. \ref{fig_HOSIDFs_Clegg} with the x-axis representing input signal frequency. Hence the corresponding point on the $H_n$ line plot represents the magnitude or phase of the $n^{th}$ harmonic, i.e., $a_n$ and $\theta_{n}$ respectively in Fig. \ref{fig_HOSIDF_principle}. It can be seen that while $|H_{\text{odd }n \geq 2}(\omega)|$ is lesser than $|H_1(\omega)|\ \forall\ \omega$, $|H_1(\omega)|$ is not significantly higher and does not dominate allowing for the exclusive use of DF for analysis.

\begin{figure}\includegraphics[trim = {1.0cm 0.5cm 1.5cm 1cm}, width=1\linewidth]{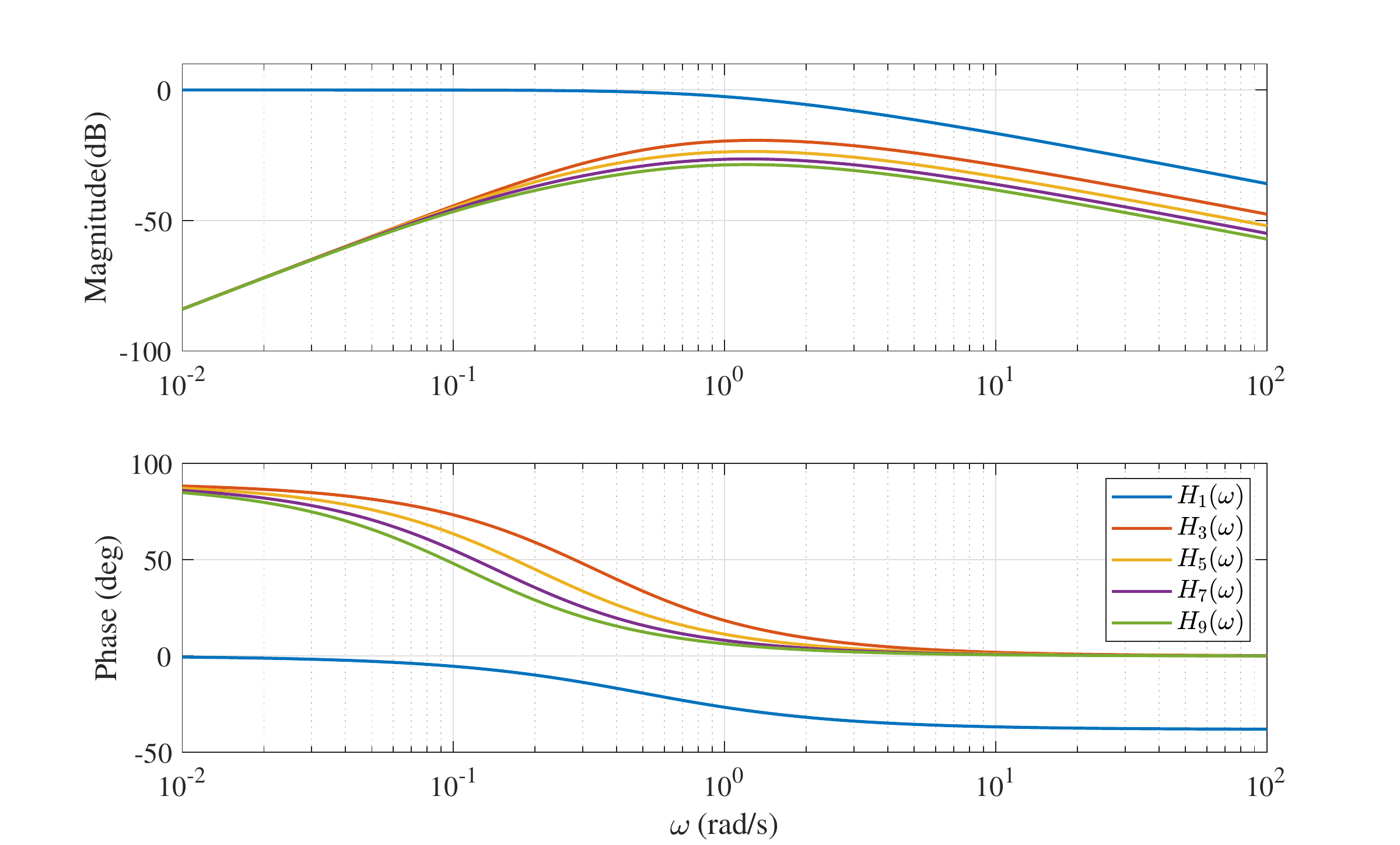}
	\centering
	
	\caption{HOSIDFs of a FORE with $\omega_r = 1$.}
	\label{fig_HOSIDFs_FORE}	
\end{figure}

Similarly, the HOSIDFs of FORE are plotted in Fig. \ref{fig_HOSIDFs_FORE}. For a FORE, at low frequencies, the phase lag between the state of FORE $x_R$ and the input $e$ is close to 0 for frequencies significantly below the cut-off $\omega_r$. Hence, the resetting action is negligible and this is seen in the low value of $|H_{\text{odd }n \geq 2}(\omega)|$. Correspondingly, for frequencies well above $\omega_r$, $|H_{\text{odd }n \geq 2}(\omega)|$ has large values and mirrors that of the Clegg integrator. For FORE, since the ratio of $|H_{\text{odd }n \geq 2}(\omega)|$ to $|H_1(\omega)|$ is not constant at all $\omega$, there must exist frequency ranges where the DF is more reliable and others where the DF is less so, especially from the context of predicting closed-loop performance.

\begin{figure}
	\centering
	\includegraphics[trim = {1.0cm 0.5cm 1.5cm 1cm}, width=1\linewidth]{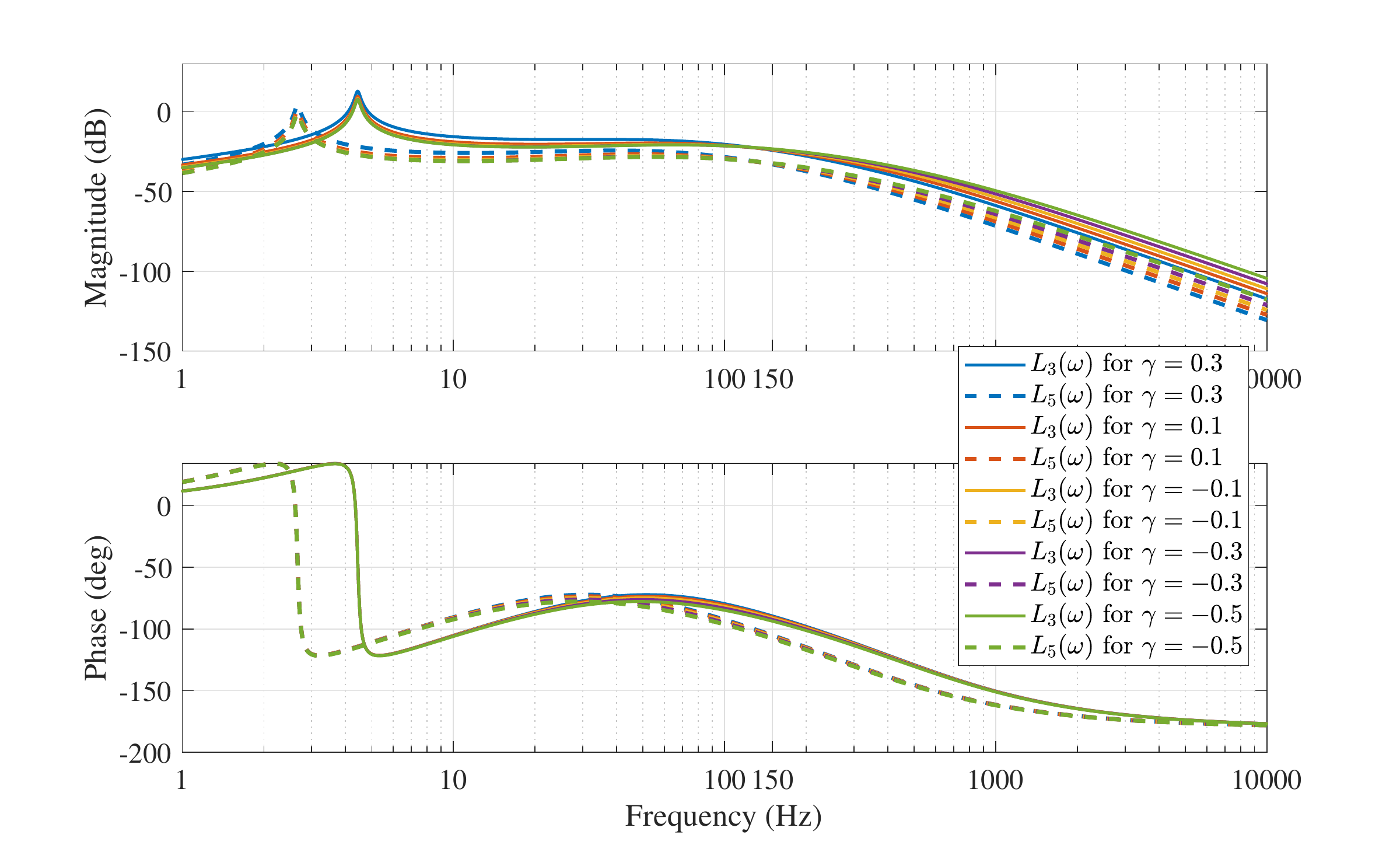}
	\caption{HOSIDFs of open-loop for $n = 3, 5$ corresponding to the DF plotted in Fig. \ref{fig_DF_OLs}.}
	\label{fig_HOSIDFs_OLs}	
\end{figure}

As a final visualization, the open-loop HOSIDFs corresponding to the DF plotted in Fig. \ref{fig_DF_OLs} are plotted for $n = 3,5$ in Fig. \ref{fig_HOSIDFs_OLs}. This shows that although the DFs were well-matched with very small differences, there is a greater difference in the HOSIDFs explaining the step response variation seen in Fig. \ref{fig_step_response_problem}. Additionally, it should be noted that due to Corollary \ref{cor_ol_hosidf}, the resonance of the plant is left-shifted in $\omega$ resulting in the $5^{th}$ harmonic dominating the $3^{rd}$ in a small range of frequencies. The HOSIDF tool provides a clear graphical visualization of the frequency response behaviour of the open-loop $\mathcal{RCS}$ and can be used to explain the difference in closed-loop behaviour of $\mathcal{RCS}$ with same or similar DF.

\section{Sensitivity functions}
\label{section_Sensitivity}

The core of loop-shaping in controller design is the relation between open-loop and closed-loop frequency behaviour. Through this, we can translate closed-loop requirements such as good reference tracking and disturbance rejection to high open-loop gain and noise rejection to low open-loop gain. Additionally, Nyquist plots allow for stability analysis. While no literature can be found for frequency domain based stability analysis of $\mathcal{RCS}$, the lack of sensitivity functions to go from open-loop to closed-loop even when stability is guaranteed hinders the use of loop-shaping with reset control. In this section, with clearly noted assumptions, we model the $\mathcal{RCS}$ such that DF and HOSIDFs can be used to predict the closed-loop behaviour and in essence allow us to translate open-loop DF and HOSIDFs to closed-loop DF and HOSIDFs.

\subsection{$\mathcal{RCS}$ with virtual harmonic generator and separator}
\label{subsection_RCS_generator_separator}

\begin{figure*} 
	\centering
	\includegraphics[trim = {0 0 0 0}, width=1\linewidth]{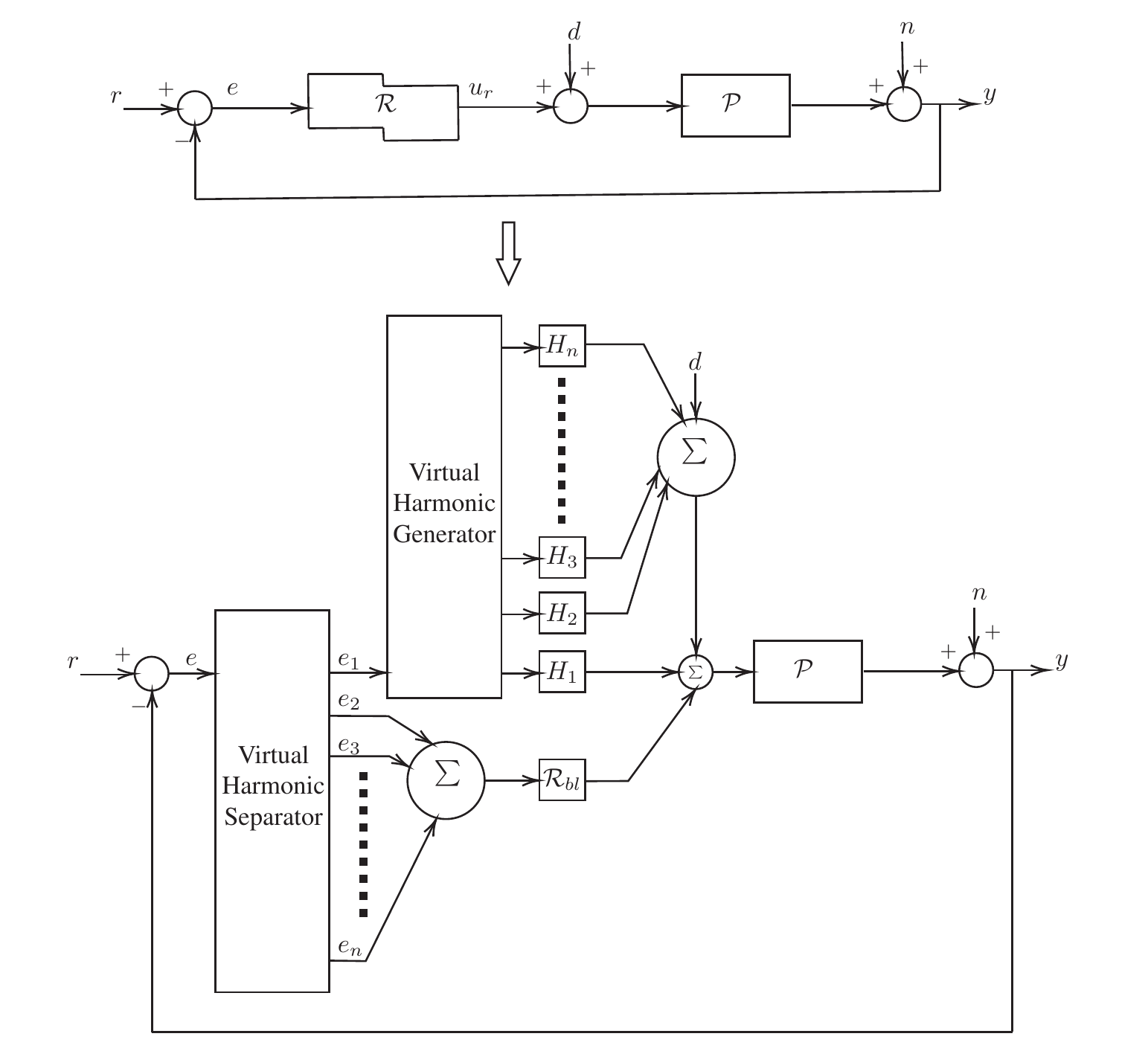}
	\caption{Representation of Higher-order sinusoidal-input describing function of reset controller $\mathcal{R}$ in closed-loop.} 
	\label{fig_HOSIDF_closed_loop}
\end{figure*}

We start with the modelling of the $\mathcal{RCS}$ of Fig. \ref{fig_RCS} to include the virtual harmonic generator as shown in Fig. \ref{fig_HOSIDF_closed_loop} to enable the inclusion of the HOSIDFs developed in the previous section in predicting the response of $\mathcal{RCS}$ to external inputs $r$, $d$ or $n$. It is clear that since each harmonic of $e$ could potentially result in multiple additional harmonics, a straight-forward assessment is cumbersome and potentially impossible. Hence, through some assumptions noted next, we simplify the closed-loop model.

\textbf{\textit{Assumption 1:}} $\mathcal{RCS}$ is input-to-state convergent.\\
\noindent $\mathcal{RCS}$ is assumed to be convergent in the sense defined in \cite{pavlov2005convergent} for the purpose of output prediction. In our previous works, we have provided results from practice which indicates that this is true. Additionally, \cite{beker2004fundamental} provides conditions for BIBO stability and \cite{beker2000forced} provides conditions under which a sinusoidal input excitation results in a periodic response. Further, the local stability of this condition is proven in \cite{beker2002analysis} with additional comments about global stability. However, currently, no mathematical proof for the same can be found in literature. Since the new sensitivity functions are developed to provide a more accurate prediction of the response and for improved controller design techniques, we consider this a reasonable assumption.

With this assumption, now in Fig. \ref{fig_HOSIDF_closed_loop}, for any sinusoidal input excitation, according to \cite{pavlov2005convergent}, $y$, $e$ and $u_R$ are periodic with the same fundamental frequency as that of the excitation. Hence, similar to what we showed in \ref{section_Hosidf}, they can be written as the summation of harmonics as below. Since a sinusoidal input $\sin(\omega t)$ is an odd function, the even harmonics in the output are also zero. 
\begin{align}
	y(t) = \sum_{n = 1}^{\infty} |Y_n| \sin(n\omega t + \angle{Y_n})\label{eq_y}\\
	e(t) = \sum_{n = 1}^{\infty} |E_n| \sin(n\omega t + \angle{E_n})\label{eq_e}\\
	u_R(t) = \sum_{n = 1}^{\infty} |U_n| \sin(n\omega t + \angle{U_n})\label{eq_u}
\end{align}
We additionally define each harmonic in the form $y_n(t) = |Y_n| \sin(n\omega t + \angle{Y_n})$. As seen above, from here on, uppercase letters are used to indicate the frequency-domain components, while lowercases are used for time-domain as per convention.

\textbf{\textit{Assumption 2:}} Reset times $t_k$ occur $\pi/\omega$ apart and result in two resets per time period.\\
\noindent If $e$ is represented as above, it can cross the zero line multiple times in a single time period of the sine wave ($2\pi/\omega$). Additionally, from the results provided in \cite{beker2002analysis} and our previous works, we know that this assumption is not true. However, we make this assumption for the following reason. In \cite{beker2002analysis}, conditions to achieve periodic output is provided which shows that in the case of multiple resets (more than 2), the interval between successive resets is not constant. Additionally, the DF used in $\mathcal{RCS}$ analysis till date and HOSIDFs developed in \ref{section_Hosidf} rely on two reset instants. Hence, while we note that this assumption can result in errors in prediction, it is necessary for the utilization of open-loop DF and HOSIDFs for prediction. 

\textbf{\textit{Assumption 3:}} Only the first harmonic of error $e$ ($e_1$) results in resets and hence the creation of higher-order harmonics ($n > 1$) in $u_R$.
\noindent Since DF and HOSIDFs are developed for a single sinusoidal excitation, we assume that $(|E_n| \forall \text{ odd }n > 1) << |E_1|$. We again note that this assumption results in errors, but are unavoidable for DF and HOSIDF based simple prediction methods. To accommodate this assumption within the $\mathcal{RCS}$ model, we introduce the concept of a virtual harmonic separator which exclusively allows passage of only the first harmonic to create resets. In essence, it behaves like a high-order anti-notch filter.

With the above assumptions, $\mathcal{RCS}$ is modelled as in Fig. \ref{fig_HOSIDF_closed_loop} to include both the virtual harmonic generator and the newly introduced virtual harmonic separator. The following conclusions can be drawn for a single sinusoidal excitation input.
\begin{enumerate}
	\item The virtual harmonic generator creates higher-order harmonics exclusively for $e_1$. The virtual harmonic separator ensures that only $e_1$ enters the virtual harmonic generator.
	\item The harmonics generated for $e_1$ are passed through the parallel interconnection of blocks $H_1$ to $H_n$ as in \ref{section_Hosidf}.
	\item The DF ($n = 1$) behaviour of $\mathcal{R}$ is desired, while the higher-order harmonics and their effects are undesired. Hence the output of blocks $H_n (\forall\ n > 1)$ are modelled as disturbances entering the system.
	\item The virtual harmonic separator ensures that the higher-order harmonics of $e$ do not influence the resetting action. Hence these harmonics are influenced by the base-linear system of $\mathcal{R}$ and not by any of the blocks $H_1$ to $H_n$. This is represented as $\mathcal{R}_{bl}$ in Fig. \ref{fig_HOSIDF_closed_loop}. $\mathcal{R}_{bl}$ can be represented by (\ref{eq_reset_controller}) without the second line (jump equation).
\end{enumerate}

The use of the virtual harmonic generator along with the virtual harmonic separator creates exclusive paths with linear blocks for the transmission of harmonic signals through the closed-loop system and enables through simplification; an easier analysis of each harmonic individually.

\subsection{Open-loop to closed-loop}
\label{subsection_ol_to_cl}

With the assumptions and the closed-loop HOSIDF representation of Fig. \ref{fig_HOSIDF_closed_loop}, the sensitivity functions to go from open-loop to closed-loop for $\mathcal{RCS}$ can be developed. We define the following notations for convenience.
\begin{align}
	L_n(\omega) = H_n(\omega)P(n\omega) \label{eq_loopol_n} \\ 
	Sl_n(\omega) = \frac{1}{1 + L_n(\omega)} \label{eq_senseol_n} \\
	L_{bl}(\omega) = \mathcal{R}_{bl}(\omega)P(\omega) \label{eq_loopol_bl} \\
	Sl_{bl}(\omega) = \frac{1}{1 + L_{bl}(\omega)} \label{eq_senseol_bl}
\end{align}

\begin{theorem}
	\label{th_ol_to_cl_r}
	With Assumptions 1 - 3, the sensitivity S ($r$ to $e$), complementary sensitivity T ($r$ to $y$) and control sensitivity CS ($r$ to $u_R$) DF and HOSIDFs can be provided as below
	\begin{align}
		S_1(\omega) & = \frac{E_1(\omega)}{R(\omega)} = Sl_1(\omega)\label{eq_sensi_1}\\
		S_{n > 1}(\omega) & = \frac{E_n(\omega)}{R(\omega)} \nonumber\\  & = -L_n(\omega)Sl_{bl}(n\omega)(|S_1(\omega)|\angle(n\angle{S_1(\omega)}))\label{eq_sensi_n}\\
		T_1(\omega) & = \frac{Y_1(\omega)}{R(\omega)} = L_1(\omega)Sl_1(\omega)\label{eq_com_sensi_1}\\
		T_{n > 1}(\omega) & = \frac{Y_n(\omega)}{R(\omega)} \nonumber\\  & = L_n(\omega)Sl_{bl}(n\omega)(|S_1(\omega)|\angle(n\angle{S_1(\omega)}))\label{eq_com_sensi_n}\\
		CS_1(\omega) & = \frac{U_1(\omega)}{R(\omega)} = H_1(\omega)Sl_1(\omega)\label{eq_con_sensi_1}\\
		CS_{n > 1}(\omega) & = \frac{U_n(\omega)}{R(\omega)} \nonumber\\  & = H_n(\omega)(1-L_{bl}(n\omega)Sl_{bl}(n\omega))\nonumber\\
		& \hspace{1.5cm}\times (|S_1(\omega)|\angle(n\angle{S_1(\omega)}))\label{eq_con_sensi_n}
	\end{align}
\end{theorem}
Proof: The exclusive closed-loop path for the first harmonic includes the virtual harmonic separator, virtual harmonic generator, $H_1$ and $\mathcal{P}$. From this we get (\ref{eq_sensi_1}), (\ref{eq_com_sensi_1}) and (\ref{eq_con_sensi_1}).

From Assumption 3 and conclusions, $e_1$ results in the generation of higher-order harmonics which are modelled as disturbance. The exclusive path for each harmonic after passing $H_{n>1}$ is through the $\mathcal{P}$, virtual harmonic separator and $\mathcal{R}_{bl}$. This provides (\ref{eq_sensi_n}) and (\ref{eq_com_sensi_n}). The $n^{th}$ harmonic of $u_R$ consists of two components. The first is the output of the virtual harmonic generator which is modelled as an external disturbance. The second component is the controller output generated as a reaction to this disturbance. This results in Eqn. \ref{eq_con_sensi_n}.

In (\ref{eq_sensi_n}), (\ref{eq_com_sensi_n}) and (\ref{eq_con_sensi_n}), $(|S_1(\omega)|\angle(n\angle{S_1(\omega)}))$ term accounts for the fact that all harmonics are generated by $e_1$ according to Assumption 3 and the phase component has the factor $n$ to account for the harmonic frequency. This concludes the proof.

\begin{theorem}
	\label{th_ol_to_cl_d}
	With Assumptions 1 - 3, the sensitivity Sd ($d$ to $e$), complementary sensitivity Td ($d$ to $y$) and control sensitivity CSd ($d$ to $u_R$) DF and HOSIDFs can be provided as below
	\begin{align}
		Sd_1(\omega) & = \frac{E_1(\omega)}{D(\omega)} = -P(\omega)Sl_1(\omega)\label{eq_sensid_1}\\
		Sd_{n > 1}(\omega) & = \frac{E_n(\omega)}{D(\omega)} \nonumber\\  & = -L_n(\omega)Sl_{bl}(n\omega)(|Sd_1(\omega)|\angle(n\angle{Sd_1(\omega)}))\label{eq_sensid_n}\\
		Td_1(\omega) & = \frac{Y_1(\omega)}{D(\omega)} = P(\omega)Sl_1(\omega)\label{eq_com_sensid_1}\\
		Td_{n > 1}(\omega) & = \frac{Y_n(\omega)}{D(\omega)} \nonumber\\  & = L_n(\omega)Sl_{bl}(n\omega)(|Sd_1(\omega)|\angle(n\angle{Sd_1(\omega)}))\label{eq_com_sensid_n}\\
		CSd_1(\omega) & = \frac{U_1(\omega)}{D(\omega)} = -L_1(\omega)Sl_1(\omega)\label{eq_con_sensid_1}\\
		CSd_{n > 1}(\omega) & = \frac{U_n(\omega)}{D(\omega)} \nonumber\\  & = H_n(\omega)(1-L_{bl}(n\omega)Sl_{bl}(n\omega))\nonumber\\
		& \hspace{1.5cm}\times (|Sd_1(\omega)|\angle(n\angle{Sd_1(\omega)}))\label{eq_con_sensid_n}
	\end{align}
\end{theorem}
\begin{theorem}
	\label{th_ol_to_cl_n}
	With Assumptions 1 - 3, the sensitivity Sn ($n$ to $e$), complementary sensitivity Tn ($n$ to $y$) and control sensitivity CSn ($n$ to $u_R$) DF and HOSIDFs can be provided as below
	\begin{align}
		Sn_1(\omega) & = \frac{E_1(\omega)}{R(\omega)} = -Sl_1(\omega)\label{eq_sensin_1}\\
		Sn_{n > 1}(\omega) & = \frac{E_n(\omega)}{R(\omega)} \nonumber\\  & = -L_n(\omega)Sl_{bl}(n\omega)(|Sn_1(\omega)|\angle(n\angle{Sn_1(\omega)}))\label{eq_sensin_n}\\
		Tn_1(\omega) & = \frac{Y_1(\omega)}{R(\omega)} = Sl_1(\omega)\label{eq_com_sensin_1}\\
		Tn_{n > 1}(\omega) & = \frac{Y_n(\omega)}{R(\omega)} \nonumber\\  & = L_n(\omega)Sl_{bl}(n\omega)(|Sn_1(\omega)|\angle(n\angle{Sn_1(\omega)}))\label{eq_com_sensin_n}\\
		CSn_1(\omega) & = \frac{U_1(\omega)}{R(\omega)} = -H_1(\omega)Sl_1(\omega)\label{eq_con_sensin_1}\\
		CSn_{n > 1}(\omega) & = \frac{U_n(\omega)}{R(\omega)} \nonumber\\  & = H_n(\omega)(1-L_{bl}(n\omega)Sl_{bl}(n\omega))\nonumber\\
		& \hspace{1.5cm}\times (|Sn_1(\omega)|\angle(n\angle{Sn_1(\omega)}))\label{eq_con_sensin_n}
	\end{align}
\end{theorem}
The paths of the harmonics are as noted before. The explanation is omitted for sake of brevity. In all cases the time domain signal can be obtained from (\ref{eq_y}), (\ref{eq_e}) and (\ref{eq_u}). Current literature on reset controllers relies on the exclusive use of DF for error prediction and hence all equations in the presented theorems related to the harmonics are neglected and only the equations pertaining to the first harmonic are used. The theorems presented allow for the calculation of closed-loop DF and HOSIDFs based on open-loop DF and HOSIDFs. The time-domain signals for $y$, $e$ and $u_R$ can then be plotted using (\ref{eq_y}), (\ref{eq_e}) and (\ref{eq_u}) respectively. 

We next shortly look at the use of this simplified model to predict the response of $\mathcal{RCS}$ when the exogenous input consists of multiple sines or when multiple exogenous inputs are present.

\subsection{Prediction with superposition}
\label{subsection_superposition}
The validity of superposition for linear systems allows for an easy analysis of systems using the sensitivity functions in the presence of multiple inputs or inputs which can be represented as a sum of multiple sinusoids or both. While this is not possible with $\mathcal{RCS}$, the use of Assumption 2 and 3 can also be extended in this case to predict the error under certain additional conditions.

\begin{corollary}
	\label{cor_superposition}
	If $w_1, w_2\cdot \cdot \cdot \cdot w_n$ are external excitation signals to $\mathcal{RCS}$ with $w_i = A_i\sin(\omega_i t + \phi_i), \forall i = 1,2,\cdot \cdot \cdot n$ and $|E_{1_i}|$ are the first harmonic error magnitudes as obtained through (\ref{eq_sensi_1}) to (\ref{eq_sensin_1}), error can be predicted under the simplified model if $|E_{1_j}| << |E_{1_k}|, \forall j = 1,2,\cdot \cdot \cdot n, j \neq k$, with $w_j$ handled by $\mathcal{R}_{bl}$.
\end{corollary}
Assumptions 2 and 3 are valid for single sinusoidal signal excitation when the magnitude of error created due to harmonics $E_{n > 1}$ is small compared to $|E_1|$, hence not resulting in multiple resets and also not significantly affecting the DF and HOSIDFs. This concept can be extended to the presence of multiple external signals. If the above condition related to $E_{1_i}$ is met, then the virtual harmonic separator ensures that the exclusive closed-loop path for signals $w_j$ are through $\mathcal{R}_{bl}$. In this case, error due to $w_k$ is predicted using (\ref{eq_sensi_1}) to (\ref{eq_sensin_n}). The additional error and related signals due to $w_j$ inputs are predicted as below.
\begin{eqnarray}
	S_j(\omega) = 
	\begin{cases}
		Sl_{bl}(\omega) & w_j \text{ is part of $r$}\\
		-P(\omega)Sl_{bl}(\omega) & w_j \text{ is part of $d$}\\
		-Sl_{bl}(\omega) & w_j \text{ is part of $n$}\\
	\end{cases}
	\label{eq_sensi_superposition}
\end{eqnarray}
\begin{eqnarray}
	T_j(\omega) = 
	\begin{cases}
		L_{bl}(\omega)Sl_{bl}(\omega) & w_j \text{ is part of $r$}\\
		P(\omega)Sl_{bl}(\omega) & w_j \text{ is part of $d$}\\
		Sl_{bl}(\omega) & w_j \text{ is part of $n$}\\
	\end{cases}
	\label{eq_com_sensi_superposition}
\end{eqnarray}
\begin{eqnarray}
	CS_j(\omega) = 
	\begin{cases}
		R_{bl}(\omega)Sl_{bl}(\omega) & w_j \text{ is part of $r$}\\
		-L_{bl}(\omega)Sl_{bl}(\omega) & w_j \text{ is part of $d$}\\
		-R_{bl}(\omega)Sl_{bl}(\omega) & w_j \text{ is part of $n$}\\
	\end{cases}
	\label{eq_con_sensi_superposition}
\end{eqnarray}
Since, $w_j$ is handled by the $\mathcal{R}_{bl}$, no additional harmonics are created.

\section{Validation}
\label{section_validation}

The accuracy of the proposed method in predicting the error $e$ and control input $u_R$ for different inputs is tested in both simulation and practice in this section. For this purpose, we make use of a precision positioning setup as explained below.

\subsection{Precision positioning setup}
\label{subsection_setup}

\begin{figure}
	\centering
	\includegraphics[width= 0.65\linewidth]{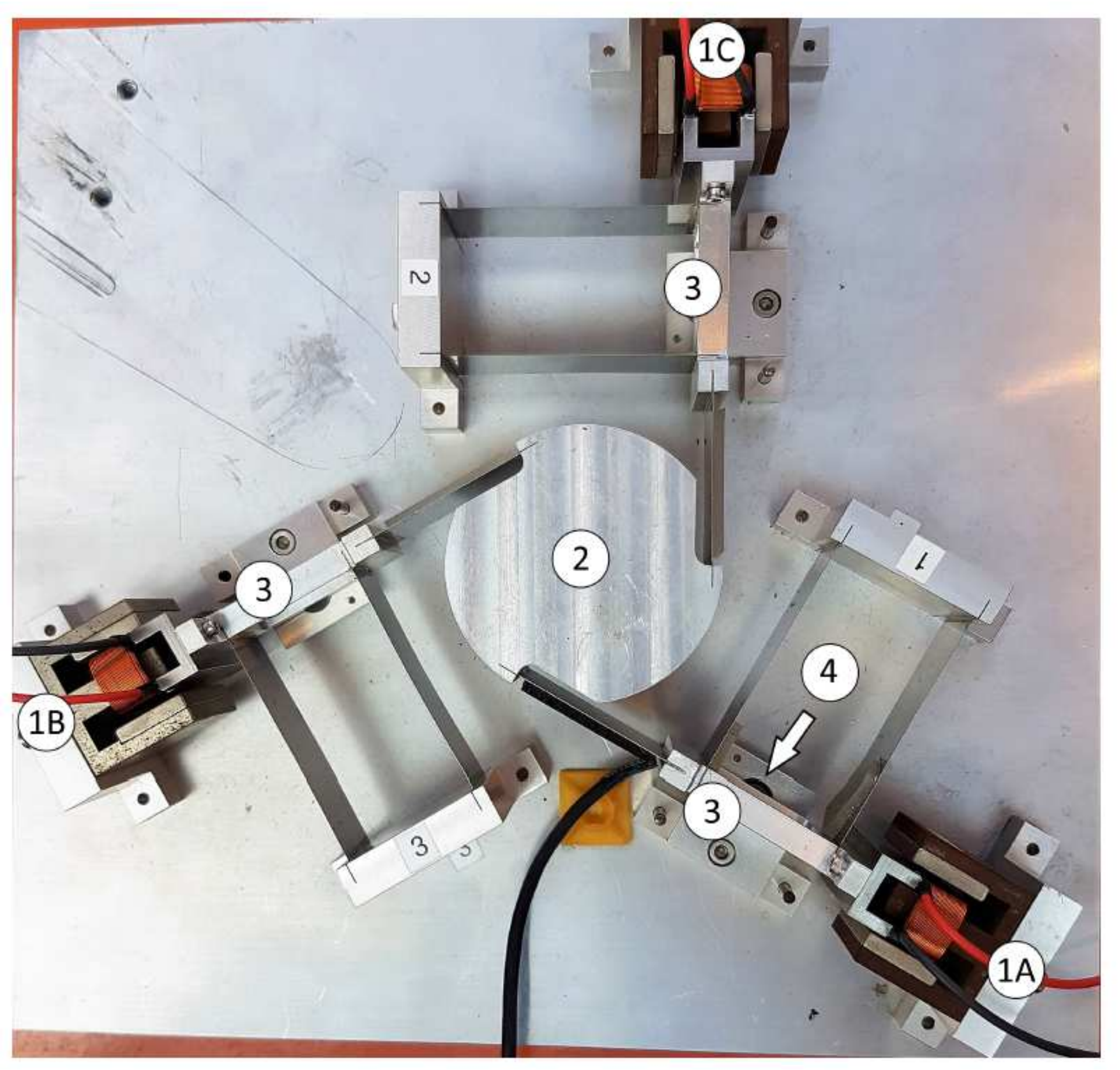}
	\caption{Planar precision positioning `Spider' stage with voice coil actuators denoted as 1A, 1B and 1C controlling the three masses (indicated as 3) and constrained by leaf flexures. The central mass (indicated by 2) is connected to these 3 masses through leaf flexures and linear encoders (indicated by 4) placed under masses `3' provide position feedback.}
	\label{fig_setup}	
\end{figure}

\begin{figure}
	\centering
	\includegraphics[trim = {1cm 0 1cm 0}, width=\linewidth]{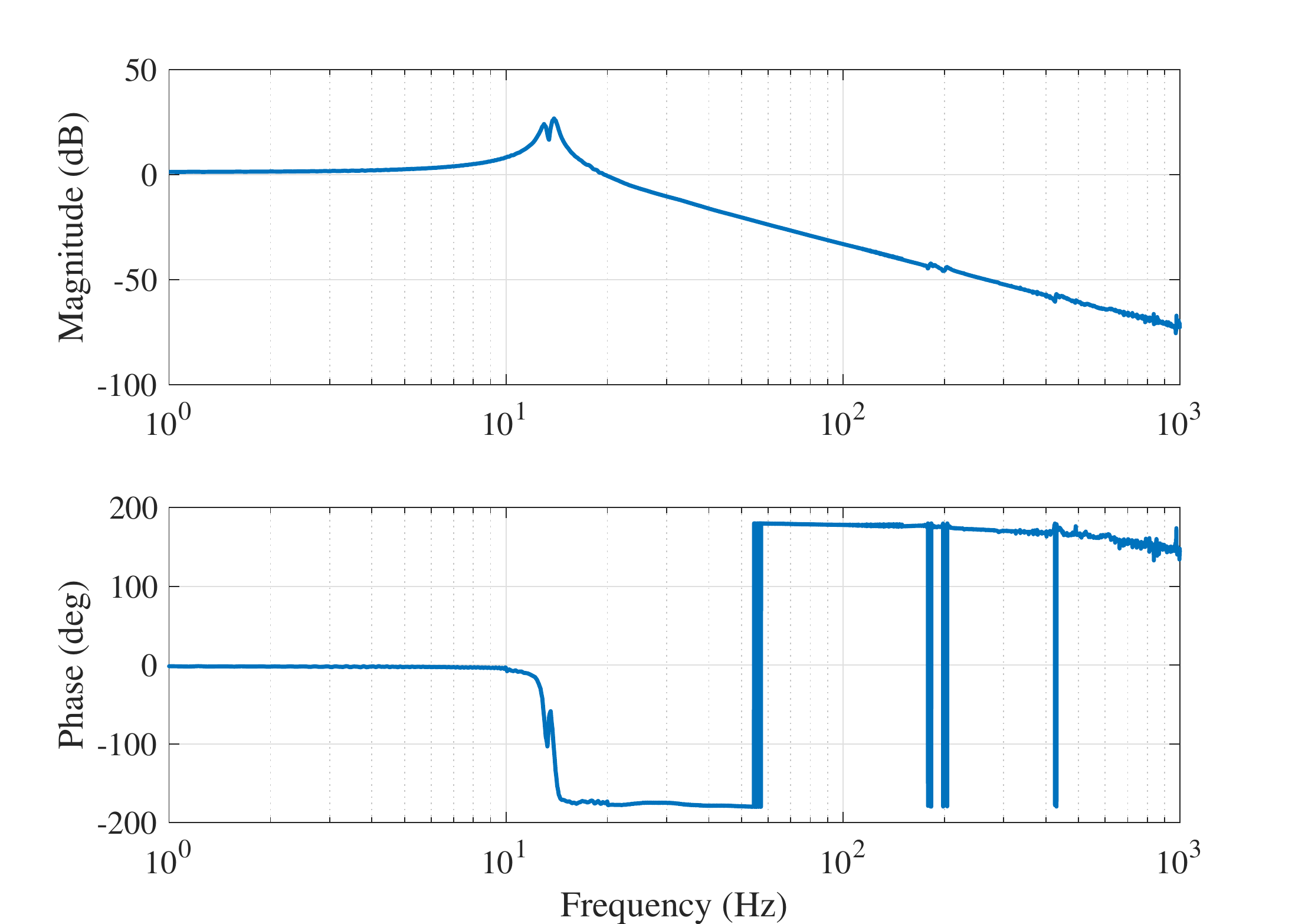}
	\caption{Frequency response data of plant as seen from actuator '1A' to position of mass '3' attached to same actuator.}
	\label{fig_frf}	
	\vspace{-2em}
\end{figure}

The precision positioning stage `Spider' shown in Fig. \ref{fig_setup} capable of planar positioning (3 DOF) is used for validation. Since reset controllers $\mathcal{R}$ is defined for SISO cases, only one of the actuators (1A) is used to position the mass `3' rigidly attached to the same. All the controllers are implemented on a NI compactRIO system with FPGA capabilities to achieve real-time control at a sampling frequency of $\SI{10}{kHz}$. Linear current source power amplifier is used to drive the voice coil actuator (1A) with a Mercury M2000 linear encoder providing position sensing with a resolution of $\SI{100}{nm}$. With additional over-sampling introduced on the FPGA, this resolution is increased to $\SI{3.125}{nm}$. The FRF of the stage is obtained as shown in Fig. \ref{fig_frf} and this shows that the plant behaviour is similar to that of a collocated double mass-spring-damper with additional dynamics at frequencies much higher that of the first resonance. In line with the industry standard, the design of controllers and prediction is carried out using this data. However, for the sake of simulation as well as stability analysis using Theorem. \ref{th_stability}, the transfer function is estimated with a single eigen mode as given in (\ref{eq_plant_spider}) (earlier used in \ref{subsection_CgLpPID_DF} to show the problem of exclusive use of DF).

\subsection{Controller designs}
\label{subsection_controller_designs}

Different controller designs with variation in the reset element used, phase lead obtained by the linear part of the controller and phase lead from the reset part are considered for validation. All controllers are designed to achieve an open-loop gain cross-over frequency ($\omega_c$) of $\SI{150}{Hz}$ ($\SI{942.48}{rad/s}$). The specifications of the various controllers are described next.

\subsubsection{Reset controllers $\mathcal{R}$ with CI}
\label{subsubsec_reset_CI}

\begin{figure}
	\centering
	\includegraphics[trim = {1.0cm 0.5cm 1.5cm 1cm}, width=1\linewidth]{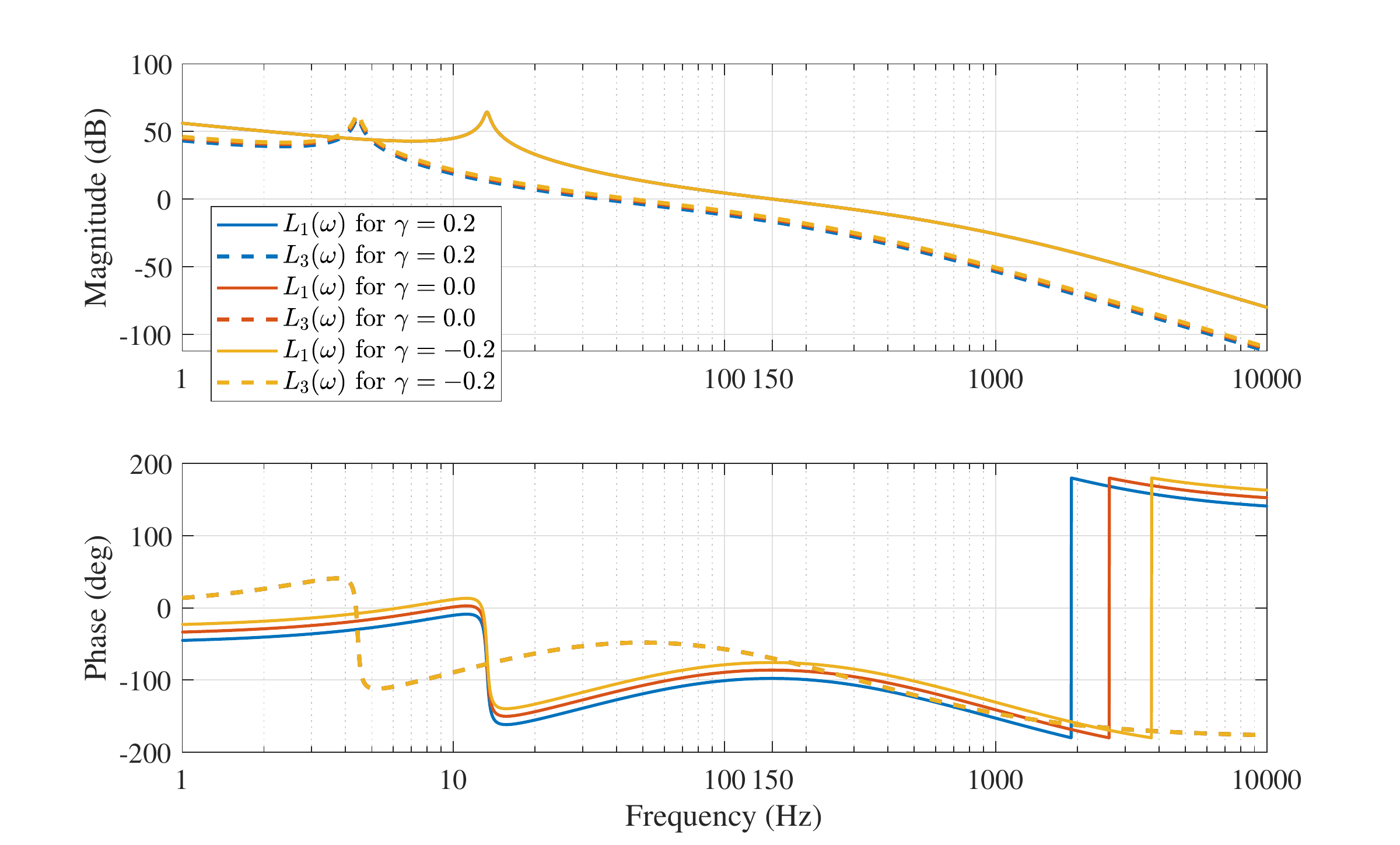}
	\caption{$L_1(\omega)$ and $L_3(\omega)$ plots for three $\mathcal{R}_{CI}$ controllers with $\gamma = \{0.2,0.0,-0.2\}$.}
	\label{fig_DF_HOSIDF_RCI}	
\end{figure}

The structure of these controllers is given below.
\begin{equation}
	\mathcal{R}_{CI} = K\underbrace{\Bigg(\frac{1}{\cancelto{\gamma}{\alpha s}}\Bigg)}_\text{Reset}\hspace{0.05cm}\underbrace{\Bigg(\frac{s + \omega_i}{\frac{s}{\omega_f} + 1}\Bigg)\Bigg(\frac{\frac{s}{\omega_d} + 1}{\frac{s}{\omega_t} + 1}\Bigg)}_\text{Non-reset}
\end{equation}
Three controllers are designed with same values of $\omega_i = \SI{15}{Hz}$, $\omega_d = \SI{50}{Hz}$, $\omega_t = \SI{450}{Hz}$ and $\omega_f = \SI{1500}{Hz}$. The difference between the controllers is in the chosen value of $\gamma = \{0.2, 0.0, -0.2\}$. The value of $K$ is corrected to ensure that DF of open-loop has a cross-over of $\omega_c = \SI{150}{Hz}$. $L_1(\omega)$ and $L_3(\omega)$ plots shown in Fig. \ref{fig_DF_HOSIDF_RCI} indicate that the change in $\gamma$ value results in a change in PM as well as $|L_3|$. It should also be noticed that in the $4 - \SI{5}{Hz}$ range, $|L_3| > |L_1|$ ensuring that pure DF based analysis cannot be carried out.

\subsubsection{Reset controllers $\mathcal{R}$ with PCI}
\label{subsubsec_reset_PCI}

\begin{figure}
	\centering
	\includegraphics[trim = {1.0cm 0.5cm 1.5cm 1cm}, width=1\linewidth]{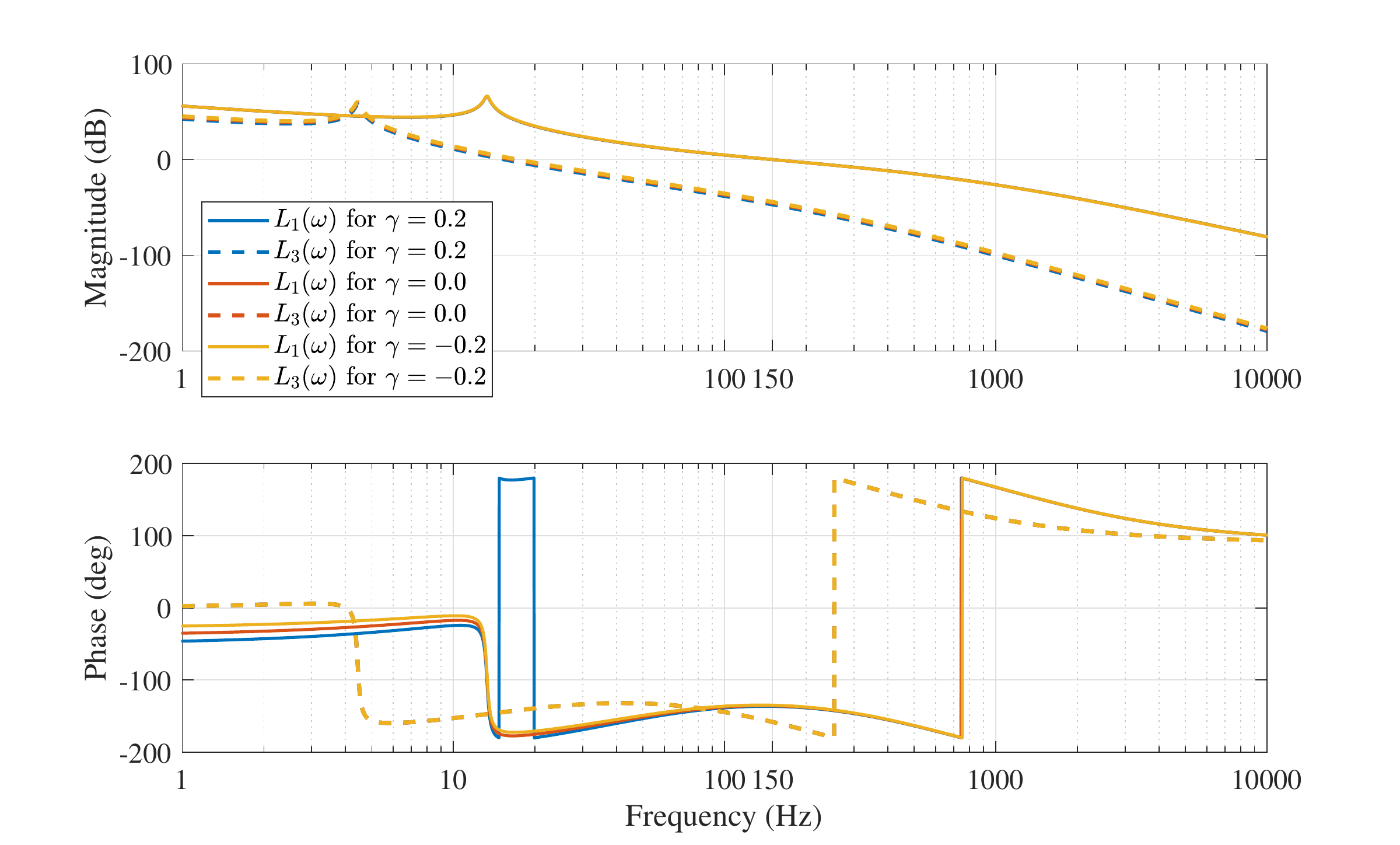}
	\caption{$L_1(\omega)$ and $L_3(\omega)$ plots for three $\mathcal{R}_{PCI}$ controllers with $\gamma = \{0.2,0.0,-0.2\}$.}
	\label{fig_DF_HOSIDF_RPCI}	
\end{figure}

While in the previous case, $1/s$ integrator is in the resetting part of $\mathcal{R}$, in this case, the complete PI filter is included in the resetting part.
\begin{equation}
	\mathcal{R}_{PCI} = K\underbrace{\Bigg(\cancelto{\gamma}{\frac{s + \omega_i}{\alpha s}}\Bigg)}_\text{Reset}\hspace{0.1cm}\underbrace{\Bigg(\frac{1}{\frac{s}{\omega_f} + 1}\Bigg)\Bigg(\frac{\frac{s}{\omega_d} + 1}{\frac{s}{\omega_t} + 1}\Bigg)}_\text{Non-reset}
\end{equation}
Three controllers are again designed with the same values provided as in the case of $\mathcal{R}_{CI}$ with their $L_1(\omega)$ and $L_3(\omega)$ plots shown in Fig. \ref{fig_DF_HOSIDF_RPCI}.

\subsubsection{CgLp-PID Reset controllers}
\label{subsubsec_reset_CgLp}

The case of the CgLp-PID controllers is unique in the sense that the CgLp element can provide phase lead ($\phi_{CgLp}$) with minimal changes to the gain behaviour in DF as seen in \ref{subsection_CgLpPID_DF}. The structure of these controllers for design using FORE is given below.
\begin{equation}
	\mathcal{R}_{CgLp} = K\underbrace{\Bigg(\frac{1}{\cancelto{\gamma}{\frac{s}{\alpha\omega_r} + 1}}\Bigg)}_\text{Reset}\hspace{0.25cm}\underbrace{\Bigg(\frac{\frac{s}{\omega_r} + 1}{\frac{s}{\omega_f} + 1}\Bigg)\Bigg(\frac{s + \omega_i}{s}\Bigg)\Bigg(\frac{\frac{s}{\omega_d} + 1}{\frac{s}{\omega_t} + 1}\Bigg)}_\text{Non-reset}
\end{equation}
As noted in \ref{subsection_CgLpPID_DF}, since CgLp-PID controllers provide a large number of tuning values with which the same $L_1(\omega)$ and PM can be achieved, a number of different $\mathcal{R}_{CgLp}$ controllers with changes in the value of $\gamma$, PM, $\phi_{CgLp}$ are designed for validation as well as an analysis of the prediction errors. The details of the designed controllers are provided in Table. \ref{tab_CgLp_controllers}.

\begin{table}
	\centering
	\begin{tabular}{|c|c|c|c|c|c|c|c|}
		\hline
		$\mathcal{R}_{CgLp}$ & PM & $\phi_{CgLp}$ & $\gamma$ & $\omega_r$ & $\alpha$ & $\omega_d$ & $\omega_t$ \\
		& ($^\circ$) & ($^\circ$) & & ($\SI{}{Hz}$) & & ($\SI{}{Hz}$) & ($\SI{}{Hz}$) \\ \hline \hline
		$\mathcal{C}_{01}$ & 50 & 30 & 0.0 & 76.08 & 1.27 & 80.17 & 280.65 \\ \hline
		$\mathcal{C}_{02}$ & \multirow{5}{*}{50} & \multirow{5}{*}{20} & 0.2 & 98.93 & 1.12 & \multirow{5}{*}{64.05} & \multirow{5}{*}{351.27} \\ \cline{1-1} \cline{4-6}
		$\mathcal{C}_{03}$ & & & 0.1 & 114.83 & 1.14 & & \\ \cline{1-1} \cline{4-6}
		$\mathcal{C}_{04}$ & & & 0.0 & 129.24 & 1.16 & & \\ \cline{1-1} \cline{4-6}
		$\mathcal{C}_{05}$ & & & -0.1 & 142.64 & 1.18 & & \\ \cline{1-1} \cline{4-6}
		$\mathcal{C}_{06}$ & & & -0.2 & 153.33 & 1.21 & & \\ \hline
		$\mathcal{C}_{07}$ & 50 & 10 & 0.0 & 230.42 & 1.07 & 49.09 & 548.29 \\ \hline
		$\mathcal{C}_{08}$ & 60 & 10 & 0.0 & 230.42 & 1.07 & \multirow{3}{*}{34.97} & \multirow{3}{*}{643.40} \\ \hline
		$\mathcal{C}_{09}$ & 70 & 20 & 0.0 & 129.24 & 1.16 & & \\ \cline{1-6}
		$\mathcal{C}_{10}$ & 80 & 30 & 0.0 & 76.08 & 1.27 & & \\ \hline
	\end{tabular}
	\caption{$\mathcal{R}_{CgLp}$ controller details with $\phi_{CgLp}$ indicating the phase lead provided by the nonlinear reset CgLp element. The common values for all the controllers are for $\omega_i = \SI{15}{Hz}$ and $\omega_f = \SI{1500}{Hz}$. $K$ is adjusted in all cases to achieve gain cross-over at $\omega_c = \SI{150}{Hz}$}
	\label{tab_CgLp_controllers}
\end{table}

\begin{figure}
	\centering
	\includegraphics[trim = {1.0cm 0.5cm 1.5cm 1cm}, width=1\linewidth]{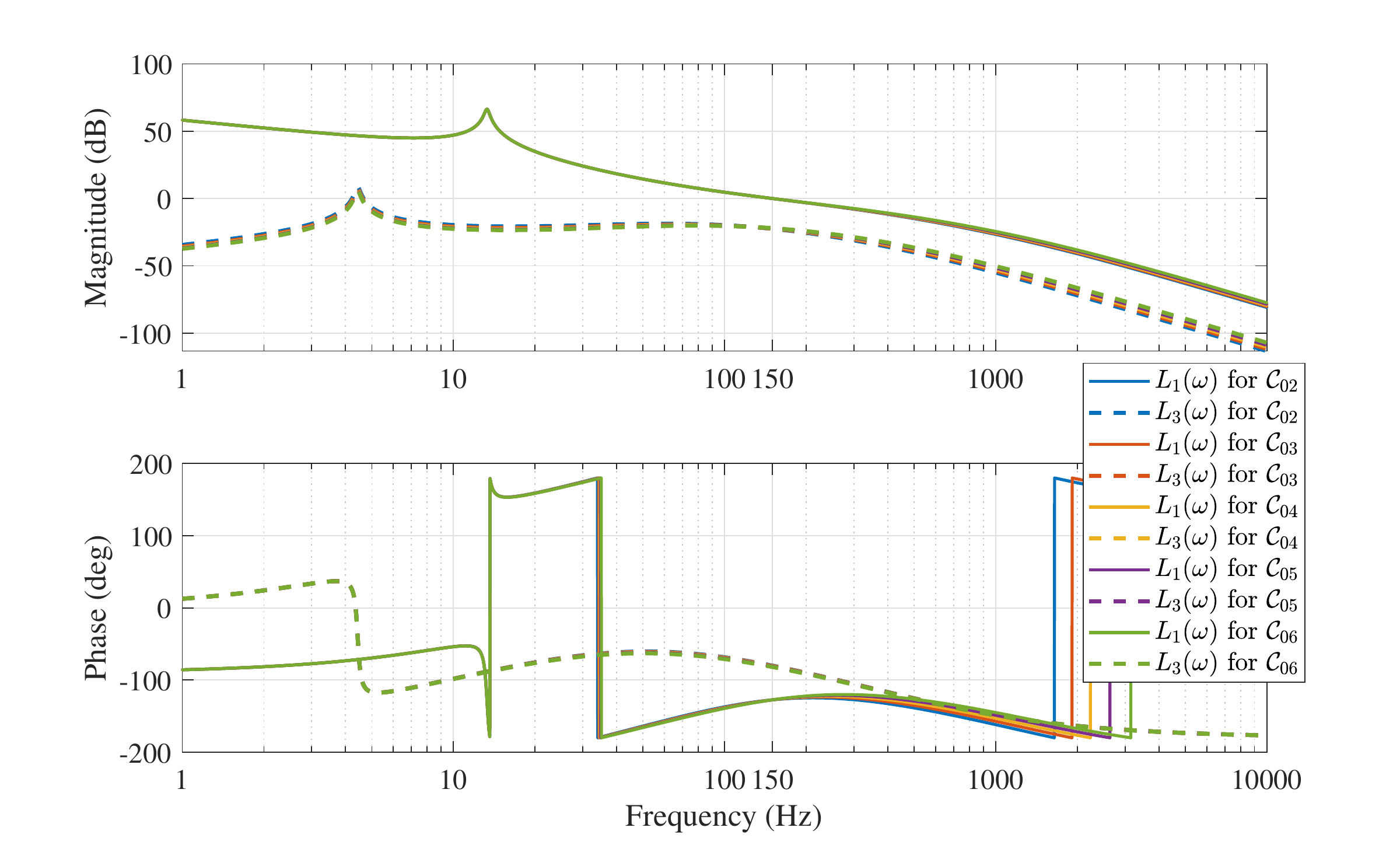}
	\caption{$L_1(\omega)$ and $L_3(\omega)$ plots for five $\mathcal{R}_{CgLp}$ controller ($\mathcal{C}_{02}$ to $\mathcal{C}_{06}$) based systems which provide same phase lead $\phi_{CgLp}$ of $20^\circ$ and same overall PM of $50^\circ$, but with different values of $\gamma$.}
	\label{fig_DF_HOSIDF_RCgLp_20}	
\end{figure}

\begin{figure}
	\centering
	\includegraphics[trim = {1.0cm 0.5cm 1.5cm 1cm}, width=1\linewidth]{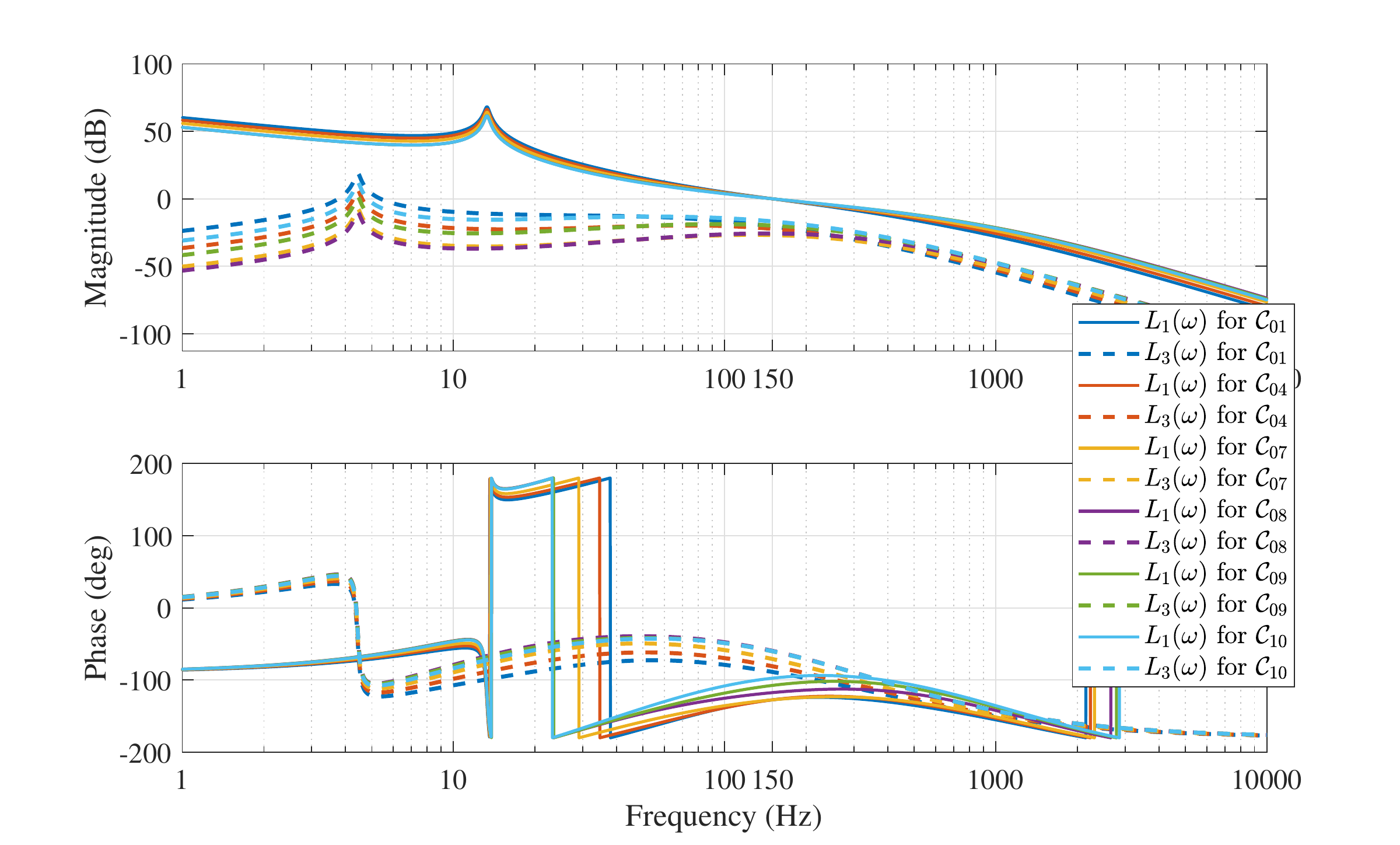}
	\caption{$L_1(\omega)$ and $L_3(\omega)$ plots for six $\mathcal{R}_{CgLp}$ controllers. $\{\mathcal{C}_{01},\mathcal{C}_{04},\mathcal{C}_{07}\}$ have same PM with different $\phi_{CgLp}$, while each group $\{\mathcal{C}_{01},\mathcal{C}_{10}\}$, $\{\mathcal{C}_{04},\mathcal{C}_{09}\}$, $\{\mathcal{C}_{07},\mathcal{C}_{08}\}$ provides same $\phi_{CgLp}$ but different PM. All controllers have same value of $\gamma = 0.0$.}
	\label{fig_DF_HOSIDF_RCgLp_diffPMphi}	
\end{figure}

$L_1(\omega)$ and $L_3(\omega)$ plots provided in Fig. \ref{fig_DF_HOSIDF_RCgLp_20} compare systems which all have same PM as well as $\phi_{CgLp}$. The different values of $\gamma$ among these controllers results in variations in $L_3$ with almost no noticeable variation in $L_1$. While the variations in $L_3$ appear small in open-loop, their effect in closed-loop can be large as seen in \ref{subsection_CgLpPID_DF}. The plots provided in Fig. \ref{fig_DF_HOSIDF_RCgLp_diffPMphi} compare systems with same value of $\gamma$, but with different PM and $\phi_{CgLp}$, resulting in variation in both $L_1$ and $L_3$.

\textbf{Comparison metrics:} Since signals $y$, $e$ and $u_R$ defined by (\ref{eq_y}), (\ref{eq_e}) and (\ref{eq_u}) for any sinusoidal excitation ($r$, $d$ or $n$) is the sum of harmonics, RMS ($L^2$ norm indicated as $||.||_2$) and maximum value ($L^\infty$ norm indicated as $||.||_\infty$) at steady-state are used as metrics to compare the prediction and measurements in both simulation and practice. The first is a popular metric used in literature and when applied to error results in integral squared average error (ISAE), while the latter is critical for precision motion control applications since the peak error determines performance in lithography applications, AFMs etc. The discontinuous nature of resetting action results in spikes in $u_R$ and can lead to saturation in many practical applications. Hence, the $L^\infty$ norm is mainly used for analysing $u_R$.

\subsection{Simulation results}
\label{subsection_simulation_results}

Simulations are run on MATLAB-Simulink for the 16 different reset controller-based $\mathcal{RCS}$ for sinusoidal excitation $r$ and $d$ with normalised amplitudes separately for a broad range of frequencies. The errors are also predicted using Theorems. \ref{th_ol_to_cl_r} and \ref{th_ol_to_cl_d}. Error is predicted in the existing literature by the exclusive use of DF and this is also calculated for comparison.

\begin{figure*}
	\centering
	\includegraphics[trim = {2.5cm 0cm 2.5cm 1cm}, width=1\linewidth]{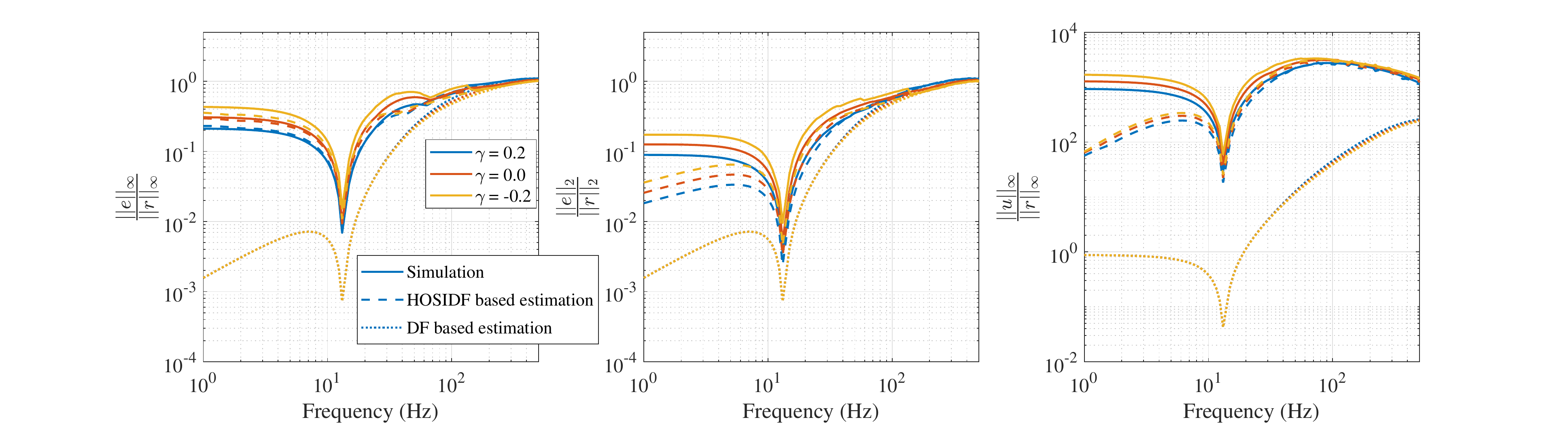}
	\caption{Sensitivity plot $\dfrac{||e||_\infty}{||r||_\infty}$, $\dfrac{||e||_2}{||r||_2}$ along with control sensitivity $\dfrac{||u_R||_\infty}{||r||_\infty}$ plotted for the three $\mathcal{R}_{CI}$ controller based $\mathcal{RCS}$.}
	\label{fig_sensitivity_RCI}    
\end{figure*}

\begin{figure*}
	\centering
	\includegraphics[trim = {2.5cm 0cm 2.5cm 1cm}, width=1\linewidth]{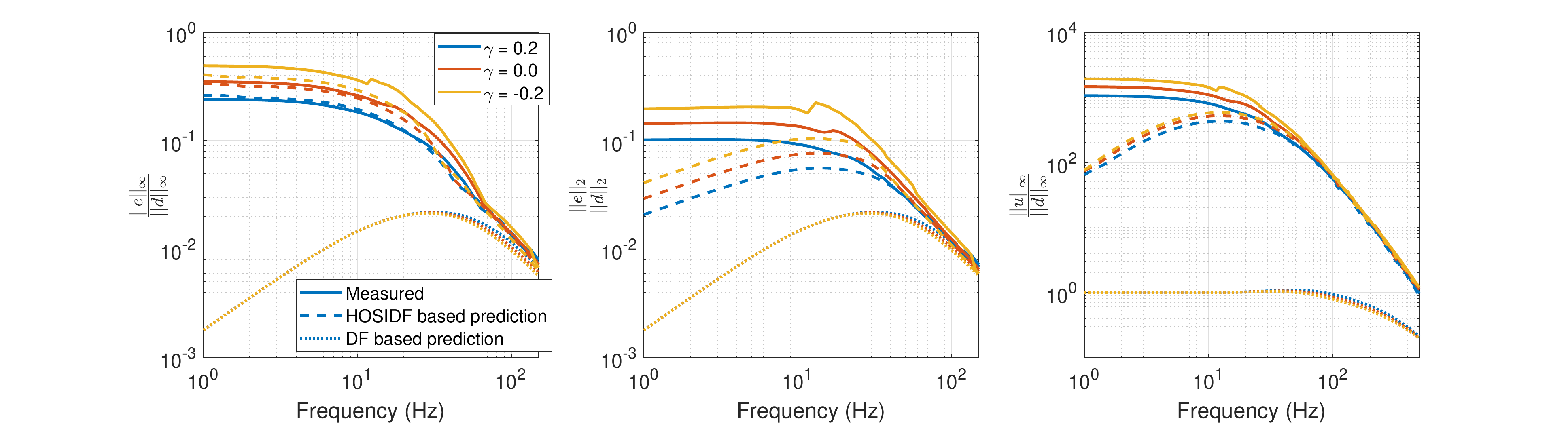}
	\caption{Process sensitivity plot $\dfrac{||e||_\infty}{||d||_\infty}$, $\dfrac{||e||_2}{||d||_2}$ along with control sensitivity to disturbance $\dfrac{||u_R||_\infty}{||d||_\infty}$ plotted for the three $\mathcal{R}_{CI}$ controller based $\mathcal{RCS}$.}
	\label{fig_process_sensitivity_RCI}    
\end{figure*}

\begin{figure*}
	\centering
	\includegraphics[trim = {2.5cm 0cm 2.5cm 1cm}, width=1\linewidth]{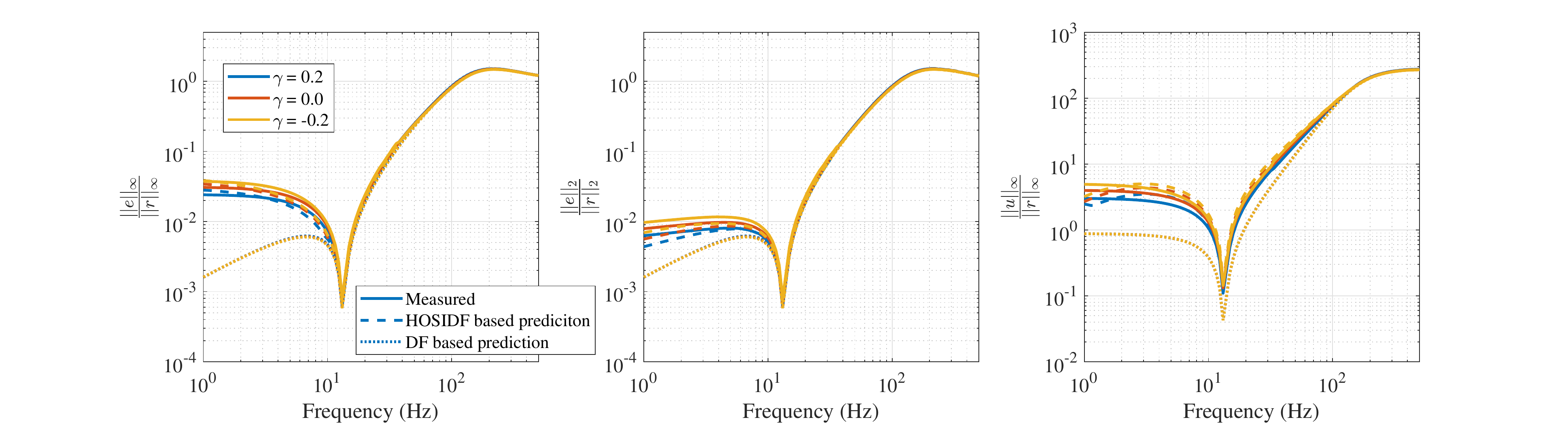}
	\caption{Sensitivity plot $\dfrac{||e||_\infty}{||r||_\infty}$, $\dfrac{||e||_2}{||r||_2}$ along with control sensitivity $\dfrac{||u_R||_\infty}{||r||_\infty}$ plotted for the three $\mathcal{R}_{PCI}$ controller based $\mathcal{RCS}$.}
	\label{fig_sensitivity_RPCI}    
\end{figure*}

\begin{figure*}
	\centering
	\includegraphics[trim = {2.5cm 0cm 2.5cm 1cm}, width=1\linewidth]{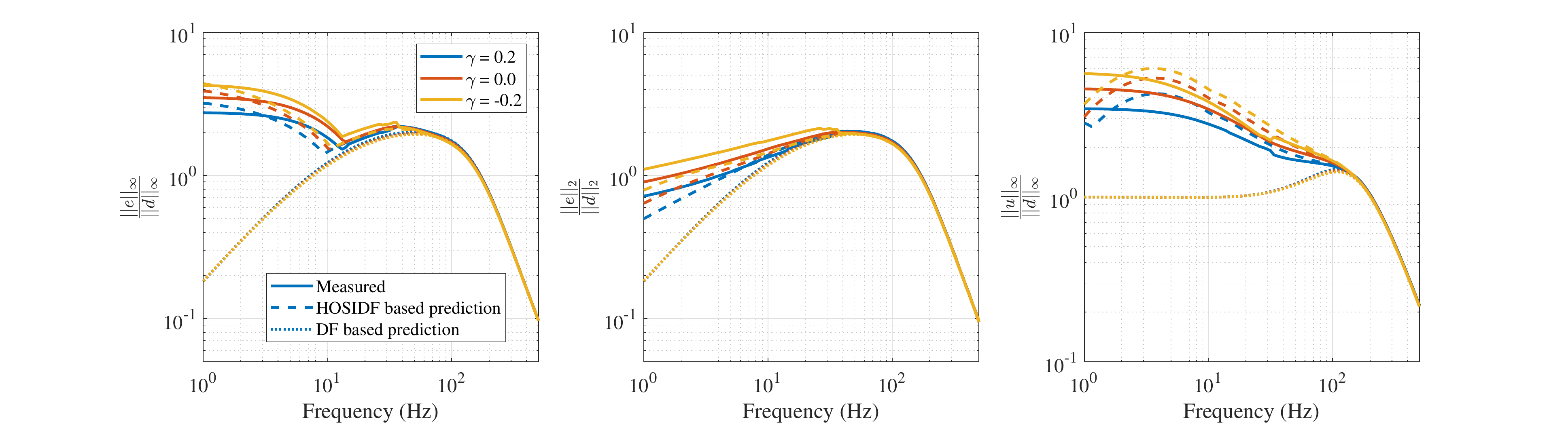}
	\caption{Process sensitivity plot $\dfrac{||e||_\infty}{||d||_\infty}$, $\dfrac{||e||_2}{||d||_2}$ along with control sensitivity to disturbance $\dfrac{||u_R||_\infty}{||d||_\infty}$ plotted for the three $\mathcal{R}_{PCI}$ controller based $\mathcal{RCS}$.}
	\label{fig_process_sensitivity_RPCI}    
\end{figure*}

The sensitivity plots created using the $L^2$ and $L^\infty$ norms of the error along with the control sensitivity plot created with the $L^\infty$ norm for input $r$ are shown in Fig. \ref{fig_sensitivity_RCI} for all three $\mathcal{R}_{CI}$ based $\mathcal{RCS}$. The same is plotted for input $d$ in Fig. \ref{fig_process_sensitivity_RCI}. These plots are also provided for all three $\mathcal{R}_{PCI}$ based $\mathcal{RCS}$ in Figs. \ref{fig_sensitivity_RPCI} and \ref{fig_process_sensitivity_RPCI}. The open-loop DF and HOSIDF open-loop plots for these systems in Figs. \ref{fig_DF_HOSIDF_RCI} and Figs. \ref{fig_DF_HOSIDF_RPCI} clearly show that the large $|L_3|$ especially with $|L_3|$ dominating $|L_1|$ in certain frequency ranges invalidates the exclusive use of DF for prediction. This is validated in the sensitivity plots where a massive difference between simulated and exclusive DF predicted values is seen. On the other hand, HOSIDF based prediction is significantly more accurate. However, we also notice that $||e||_\infty$ prediction is significantly better than that of $||e||_2$ at low frequencies. This is because resetting of the integrator results in limit cycles as noted in \cite{banos2011reset}, and hence several resets within a single time period of the sinusoidal input and a violation of both Assumption 2 and 3.

\begin{figure}
	\centering
	\includegraphics[trim = {1.0cm 0cm 1.0cm 1cm}, width=1\linewidth]{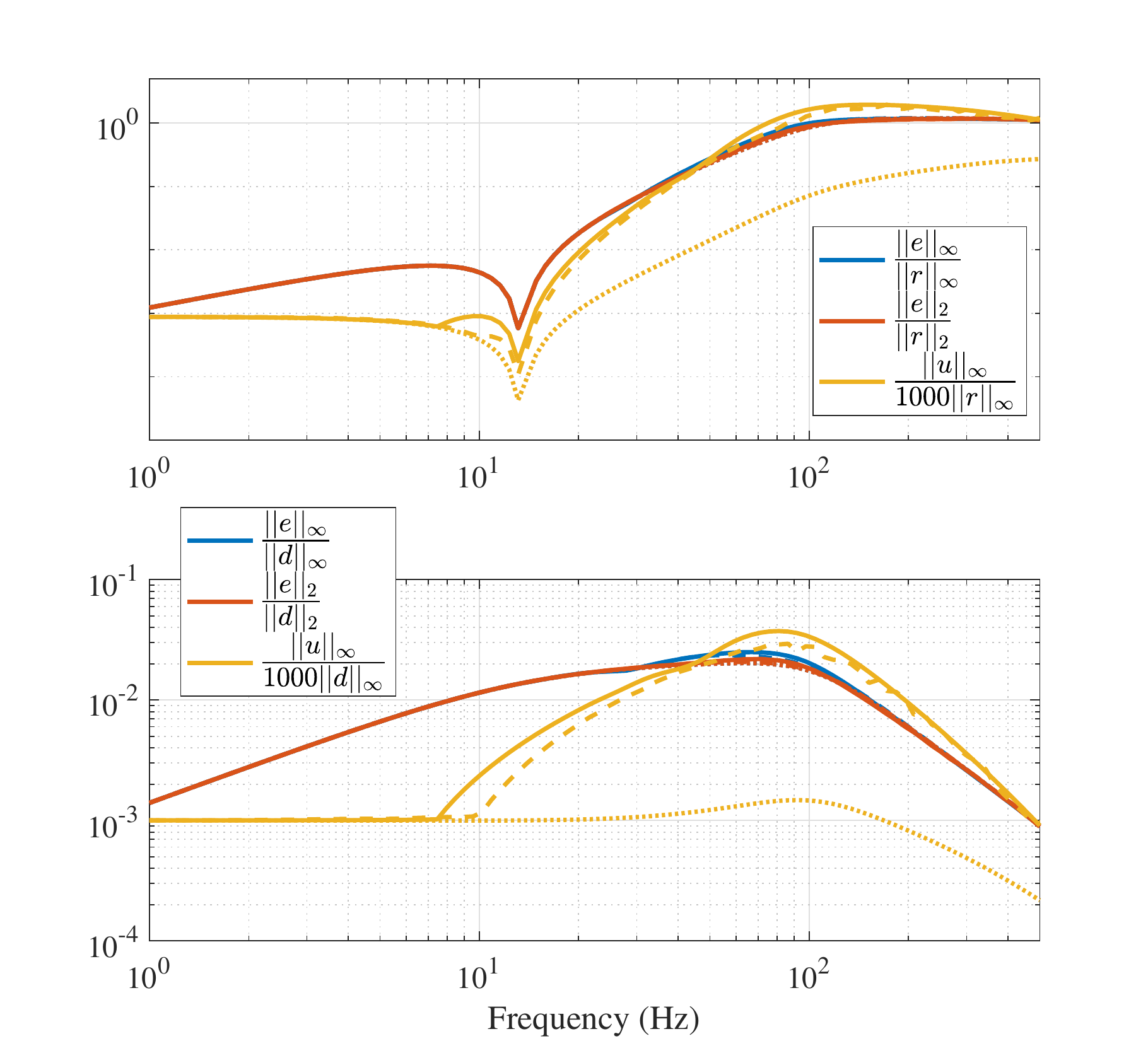}
	\caption{Sensitivity plot $\dfrac{||e||_\infty}{||r||_\infty}$, $\dfrac{||e||_2}{||r||_2}$, control sensitivity $\dfrac{||u_R||_\infty}{||r||_\infty}$, process sensitivity $\dfrac{||e||_\infty}{||d||_\infty}$, $\dfrac{||e||_2}{||d||_2}$ and control sensitivity to disturbance $\dfrac{||u_R||_\infty}{||d||_\infty}$ plotted for $\mathcal{C}_{04}$ based $\mathcal{RCS}$. Solid lines - `Simulation', Dashed lines - `HOSIDF based prediction', `Dotted lines' - DF based prediction'.}
	\label{fig_sensitivity_CCgLp_C4}    
\end{figure}

In the case of all the $\mathcal{R}_{CI}$ and $\mathcal{R}_{PCI}$ based $\mathcal{RCS}$, while the $|L_1|$ plots have almost no noticeable difference, the PM is different in all cases. Hence, now we compare the $\mathcal{R}_{CgLp}$ based $\mathcal{RCS}$ systems where the $|L_1|$ as well as PM is the same for $\mathcal{C}_{02}$ to $\mathcal{C}_{06}$ based systems. The plots as provided previously are provided for $\mathcal{C}_{04}$ in Fig. \ref{fig_sensitivity_CCgLp_C4}. In the case of $\mathcal{R}_{CgLp}$ controllers, the magnitude of the higher-order harmonics is always lower than that of the first harmonic. Hence, we can see that the DF based prediction method is also accurate in predicting the performance, especially at low frequencies. While there is a clear difference in the estimation of control input $u_R$, this is not clear in the case of error $e$. To visualise the prediction difference between the two methods, a different metric (as given below) is used.
\begin{equation}
	\text{Prediction error ratio (PER)} = \frac{|\text{Measured - Predicted}|}{\text{Predicted}}
\end{equation}

\begin{figure}
	\centering
	\includegraphics[trim = {2.6cm 0cm 5.4cm 1cm}, width=1\linewidth]{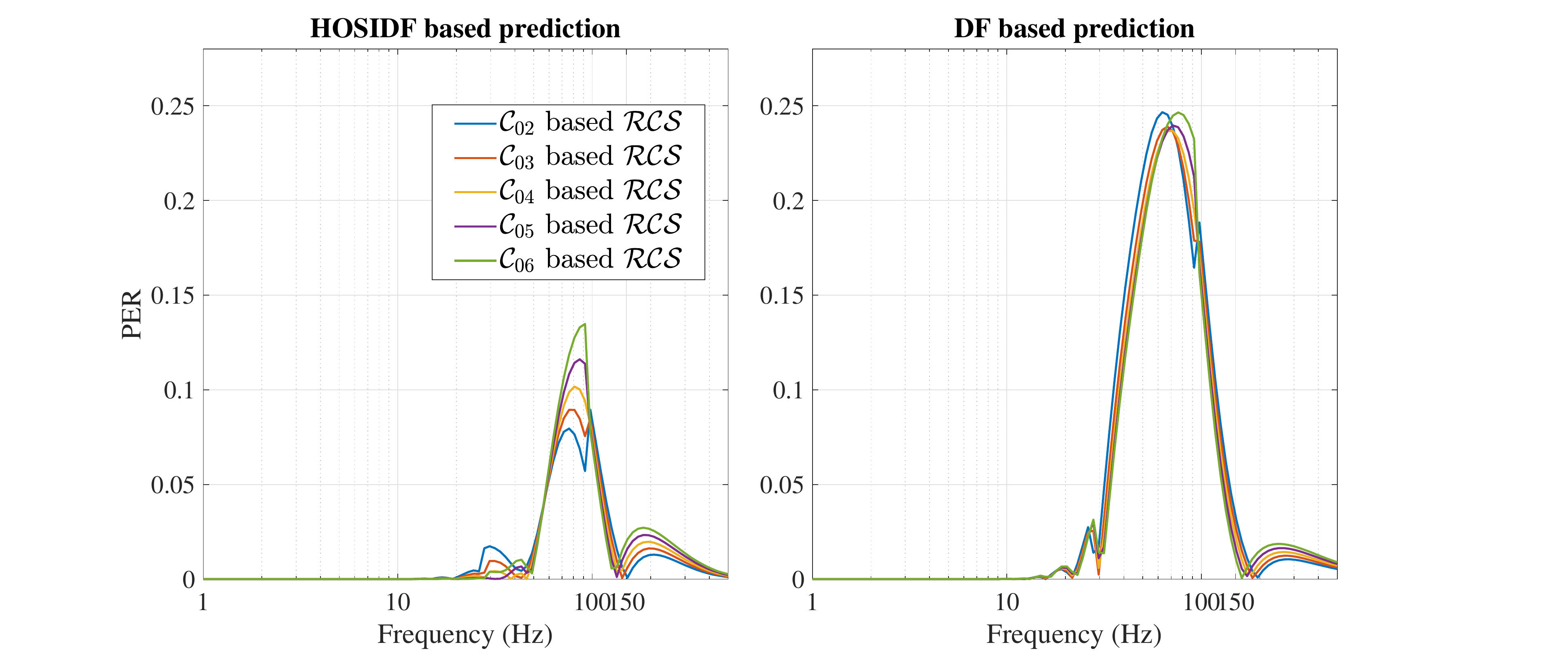}
	\caption{Prediction error ratio plots of $\mathcal{C}_{02}$ to $\mathcal{C}_{06}$ based systems for input $r$ based on $L^\infty$ norm for the existing DF based and novel HOSIDF based methods. All systems have same $|L_1|$ and PM.}
	\label{fig_sensitivity_PER_CCgLp_C2_C6}    
\end{figure}

\begin{figure}
	\centering
	\includegraphics[trim = {2.6cm 0cm 5.4cm 1cm}, width=1\linewidth]{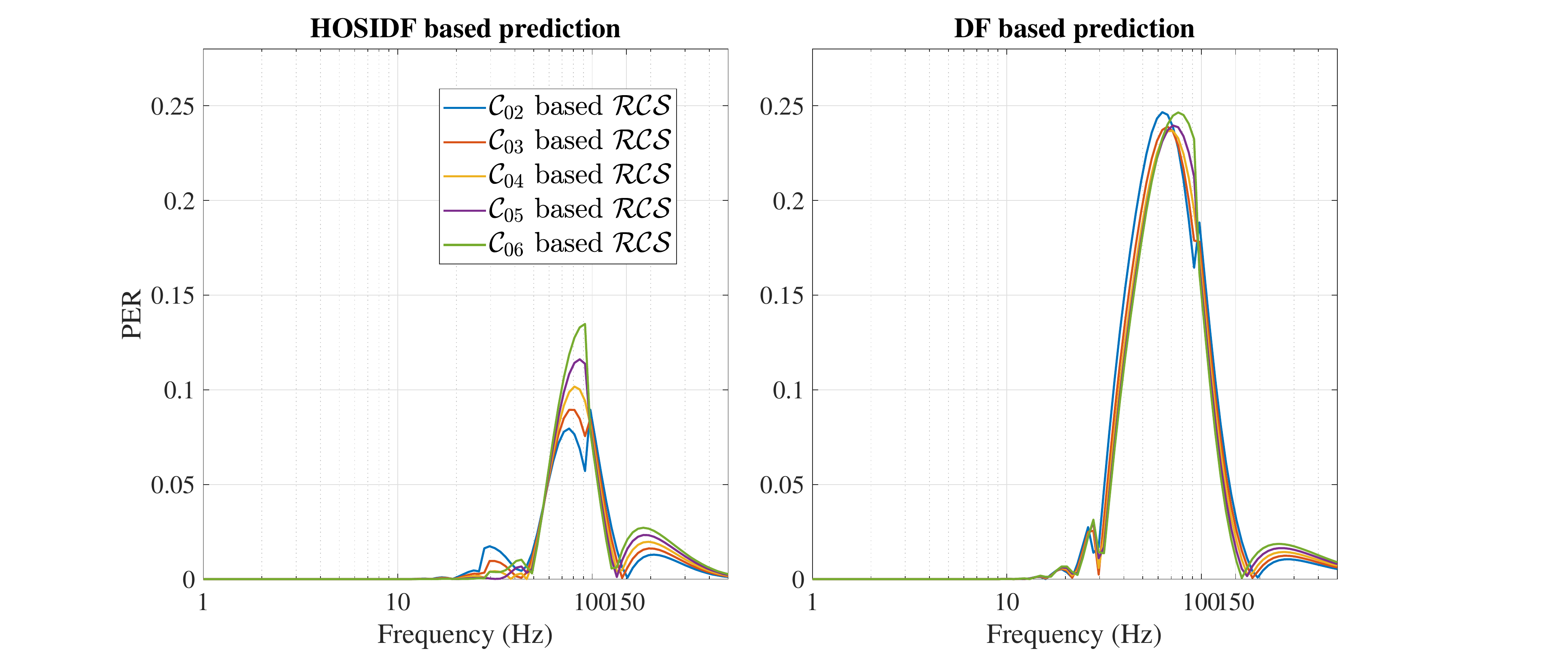}
	\caption{Prediction error ratio plots of $\mathcal{C}_{02}$ to $\mathcal{C}_{06}$ based systems for input $d$ based on $L^\infty$ norm for the existing DF based and novel HOSIDF based methods. All systems have same $|L_1|$ and PM.}
	\label{fig_process_sensitivity_PER_CCgLp_C2_C6}    
\end{figure}

\begin{figure}
	\centering
	\includegraphics[trim = {2.6cm 0cm 5.4cm 1cm}, width=1\linewidth]{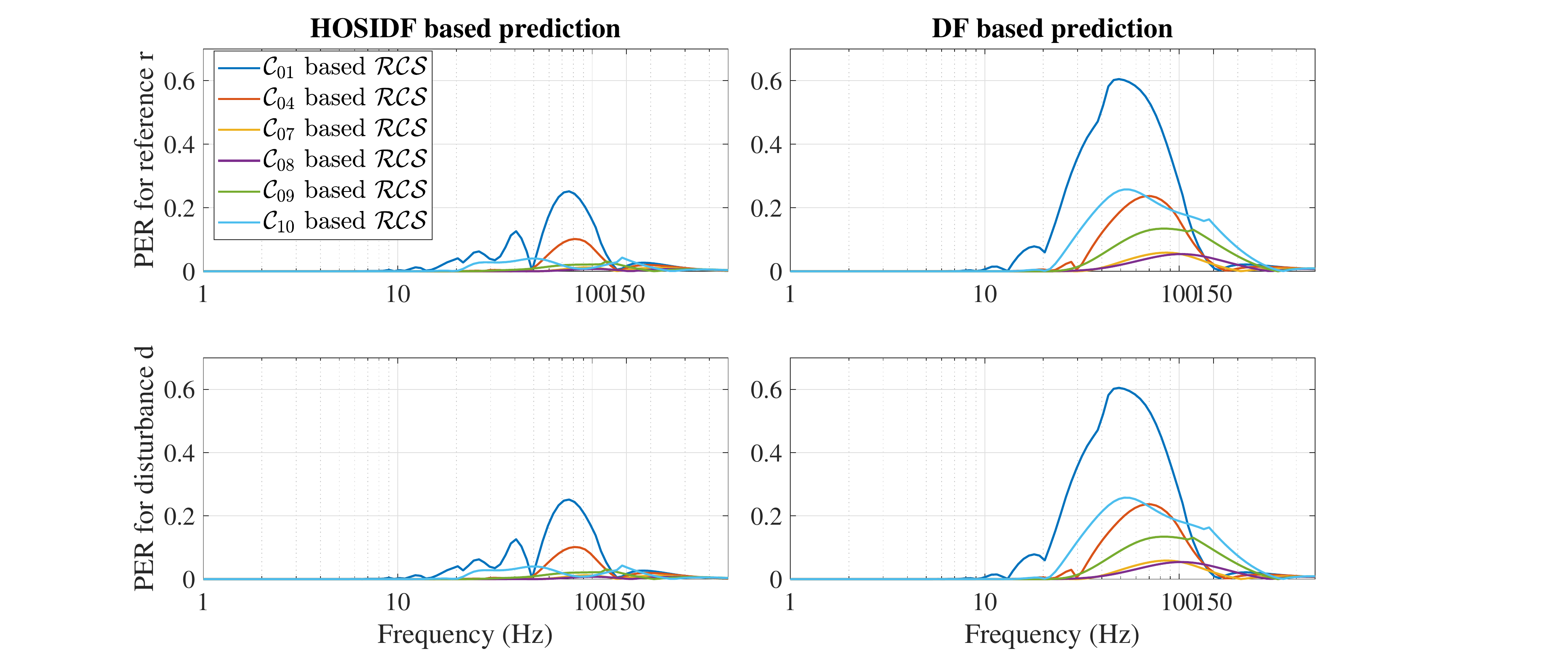}
	\caption{Prediction error ratio plots for inputs $r$ and $d$ using $L^\infty$ norm of $\mathcal{RCS}$ whose open-loop DF and HOSIDF plots are shown in Fig. \ref{fig_DF_HOSIDF_RCgLp_diffPMphi}.}
	\label{fig_sensitivity_process_sensitivity_PER_comp_CCgLp}    
\end{figure}

$\text{PER}$ plots are provided for $||e||_\infty$ for all $\mathcal{C}_{02}$ to $\mathcal{C}_{06}$ based systems in Figs. \ref{fig_sensitivity_PER_CCgLp_C2_C6} and \ref{fig_process_sensitivity_PER_CCgLp_C2_C6}. Additional plots comparing performances of the different groups of $\mathcal{R}_{CgLp}$ based $\mathcal{RCS}$ systems whose open-loop plots are given in Fig. \ref{fig_DF_HOSIDF_RCgLp_diffPMphi}, are provided in Fig. \ref{fig_sensitivity_process_sensitivity_PER_comp_CCgLp}. Plots of $||e||_2$ and $||u_R||_\infty$ are not provided for sake of brevity. These clearly show the huge difference in accuracy between the novel HOSIDF and existing DF based method.

As noted in \ref{subsection_CgLpPID_DF}, the main motivation for HOSIDF and the subsequent use of the same for error prediction is for optimal tuning. The proposed method must be capable of predicting differences in performance while the existing DF based method cannot, especially in the case of $C_{02}$ to $C_{06}$ controller based systems, although from PER plots, it is clear that Assumptions 2 and 3 leads to inaccurate prediction. This along with additional measurements from the practical setup are presented in the next subsection.

\subsection{Practical results}
\label{subsection_practical_results}

The results presented in the previous subsection are derived from simulations. The nonlinear nature of reset controllers which involves the requirement of information on the zero-crossing of the error for simulation can result in slightly different results based on the simulation settings. More importantly, practical implementation requires discretization and involves quantization of the sensed output $y$ (position in the case of Spider stage) and control input $u_R$ (voltage output of NI DAC), with the design of controllers achieved using FRF. Hence, additional results from practice are provided in this section to validate the method as well as to validate the results of simulations. Additionally as noted, the ability to predict the trend in error for different controller-based reset systems is tested. However, due to the time-consuming nature of measurements, limited results are provided in this case.

\begin{table}
	\centering
	\begin{tabular}{|c|c||c|c|c|}
		\hline
		& \backslashbox{Freq (Hz)}{$\gamma$} & $0.2$ & $0.0$ & $-0.2$ \\ \hline \hline
		Measured & \multirow{3}{*}{1 (r)} & -31.8839 & -29.8651 & -28.1686  \\ \cline{1-1} \cline{3-5}
		HOSIDF & & \multirow{2}{*}{-31.0563} & \multirow{2}{*}{-29.3309} & \multirow{2}{*}{-28.3264}  \\ 
		estimated & & & & \\ \hline \hline
		
		Measured & \multirow{3}{*}{5 (r)} & -34.0689 & -32.3537 & -31.4112  \\ \cline{1-1} \cline{3-5}
		HOSIDF & & \multirow{2}{*}{-35.6872} & \multirow{2}{*}{-34.3471} & \multirow{2}{*}{-33.7416}  \\ 
		estimated & & & & \\ \hline \hline
		
		Measured & \multirow{3}{*}{10 (r)} & -41.5930 & -40.1398 & -39.7321  \\ \cline{1-1} \cline{3-5}
		HOSIDF & & \multirow{2}{*}{-45.0022} & \multirow{2}{*}{-44.4342} & \multirow{2}{*}{-43.8006}  \\ 
		estimated & & & & \\ \hline \hline
		
		Measured & \multirow{3}{*}{1 (d)} & -30.7441 & -28.7310 & -27.0149  \\ \cline{1-1} \cline{3-5}
		HOSIDF & & \multirow{2}{*}{-29.9202} & \multirow{2}{*}{-28.1948} & \multirow{2}{*}{-27.1903}  \\ 
		estimated & & & & \\ \hline \hline
		
		Measured & \multirow{3}{*}{5 (d)} & -31.8163 & -30.0488 & -28.9385  \\ \cline{1-1} \cline{3-5}
		HOSIDF & & \multirow{2}{*}{-33.2784} & \multirow{2}{*}{-31.9383} & \multirow{2}{*}{-31.3328}  \\ 
		estimated & & & & \\ \hline \hline
		
		Measured & \multirow{3}{*}{10 (d)} & -34.0161 & -32.4751 & -31.8186  \\ \cline{1-1} \cline{3-5}
		HOSIDF & & \multirow{2}{*}{-36.7009} & \multirow{2}{*}{-36.1330} & \multirow{2}{*}{-35.4993}  \\ 
		estimated & & & & \\ \hline
	\end{tabular}
	\caption{Trends in measured and predicted $\dfrac{||e||_\infty}{||r||_\infty}$ (provided in dB) for $\mathcal{R}_{PCI}$ based $\mathcal{RCS}$ which all have the same $|L_1|$ (resulting in same error values estimated by DF) and a slight difference in PM. $r$ after the frequency indicates reference tracking, with $d$ indicating disturbance rejection.}
	\label{tab_error_trends_PCI}
\end{table}

\begin{table}
	\centering
	\begin{adjustbox}{angle=90}
		\begin{tabular}{|c|c||c|c|c|c|c|}
			\hline
			& \backslashbox{Frequency (Hz)}{Controller} & $\mathcal{C}_{02}$ & $\mathcal{C}_{03}$ & $\mathcal{C}_{04}$ & $\mathcal{C}_{05}$ & $\mathcal{C}_{06}$ \\ \hline \hline
			Measured & \multirow{3}{*}{40 (r)} & -17.3230 & -17.4016 & -17.4046 & -17.4296 & -17.3886 \\ \cline{1-1} \cline{3-7}
			HOSIDF estimated & & -16.0769 & -16.1814 & -16.2313 & -16.2482 & -16.2418 \\ \hline \hline
			
			Measured & \multirow{3}{*}{80 (r)} & -2.4686 & -2.4413 & -2.3448 & -2.2517 & -2.1694 \\ \cline{1-1} \cline{3-7}
			HOSIDF estimated & & -3.3724 & -3.3559 & -3.3407 & -3.3252 & -3.3086 \\ \hline \hline
			
			Measured & \multirow{3}{*}{90 (r)} & -0.8981 & -0.7066 & -0.4927 & -0.3170 & -0.1486 \\ \cline{1-1} \cline{3-7}
			HOSIDF estimated & & -1.9025 & -1.8473 & -1.8040 & -1.7673 & -1.7348 \\ \hline \hline
			
			Measured & \multirow{3}{*}{80 (d)} & -31.4922 & -31.4759 & -31.4759 & -31.4434 & -31.1353 \\ \cline{1-1} \cline{3-7}
			HOSIDF estimated & & -33.2106 & -33.1941 & -33.1789 & -33.1634 & -33.1468 \\ \hline \hline
			
			Measured & \multirow{3}{*}{90 (d)} & -32.2791 & -32.0045 & -31.8095 & -31.6889 & -31.3640 \\ \cline{1-1} \cline{3-7}
			HOSIDF estimated & & -33.8384 & -33.7833 & -33.7399 & -33.7033 & -33.6707 \\ \hline \hline
			
			Measured & \multirow{3}{*}{100 (d)} & -32.8365 & -32.7575 & -32.6404 & -32.4487 & -32.0411 \\ \cline{1-1} \cline{3-7}
			HOSIDF estimated & & -34.6218 & -34.5437 & -34.4827 & -34.4325 & -34.3895 \\ \hline
		\end{tabular}
	\end{adjustbox}
	\caption{Trends in measured and predicted $\dfrac{||e||_\infty}{||r||_\infty}$ (provided in dB) for $\mathcal{C}_{02}$ to $\mathcal{C}_{06}$ based $\mathcal{RCS}$ which all have the same $|L_1|$ (resulting in same error values estimated by DF) and PM. $r$ after the frequency indicates reference tracking, with $d$ indicating disturbance rejection.}
	\label{tab_error_trends}
\end{table}

The measured $\dfrac{||e||_\infty}{||r||_\infty}$ values for both reference tracking and disturbance rejection are provided in Table. \ref{tab_error_trends_PCI} for $\mathcal{R}_{PCI}$ based $\mathcal{RCS}$. Since all three systems have the same $|L_1|$ as seen in Fig. \ref{fig_DF_HOSIDF_RPCI} with small variations in the phase, the DF predicted error has very small difference between the systems. However, from Table. \ref{tab_error_trends_PCI}, large changes in the measured $\dfrac{||e||_\infty}{||r||_\infty}$ is seen. Although the novel HOSIDF based estimation does not match the measured values in all cases, the trend in $\dfrac{||e||_\infty}{||r||_\infty}$ values (increasing or decreasing with change in $\gamma$) is captured. This trend is also checked for $\mathcal{C}_{02}$ to $\mathcal{C}_{06}$ based $\mathcal{RCS}$, as these controllers provide the best overall performance. The trends are mainly checked for at frequencies where the maximum $\text{PER}$ values are seen in Figs. \ref{fig_sensitivity_PER_CCgLp_C2_C6} and \ref{fig_process_sensitivity_PER_CCgLp_C2_C6} and these values are tabulated in Table. \ref{tab_error_trends}. As expected, while the novel HOSIDF method does not completely accurately predict the error values at all frequencies, the trend in the $\dfrac{||e||_\infty}{||r||_\infty}$ values is captured which allows for a HOSIDF estimation based optimised controller tuning for these family of controllers.

\begin{table*}
	\centering
	\begin{adjustbox}{angle=90}
		\begin{tabular}{|c||c|c|c|c|c|c|c|c|c|}
			\hline
			Trial & RCS ($w_1$) & BLS ($w_1$) & RCS ($w_2$) & BLS ($w_2$) & & & & & Measured \\
			No. & $||e||_\infty$ (r) & $||e||_\infty$ (r) & $||e||_\infty$ (d) & $||e||_\infty$ (d) & & & & & $||e||_\infty$ \\ \hline
			& $X_1$ & $X_2$ & $X_3$ & $X_4$ & $X_1 + X_3$ & $X_1 + X_4$ & $X_2 + X_3$ & $X_2 + X_4$ &  \\ \hline \hline
			1 & 16.7031 & 15.5312 & 3.9531 & 3.1406 & 20.6562 & 19.5625 & 19.4843 & 18.3906 & 19.8281 \\ \hline
			2 & 16.7031 & 15.5312 & 4.5781 & 3.8437 & 21.2812 & 20.5468 & 20.1093 & 19.375 & 20.5781  \\ \hline	 
			3 & 16.7031 & 15.5312 & 8.8437 & 7.9218 & 25.5468 & 24.625 & 24.375 & 23.4531 & 24.5312 \\ \hline
			4 & 16.7031 & 15.5312 & 11.0468 & 9.9531 & 27.75 & 26.6562 & 26.5781 & 25.4843 & 27.4062 \\ \hline
			5 & 16.7031 & 15.5312 & 17.0781 & 15.1718 & 33.7812 & 31.875 & 32.6093 & 30.7031 & 35.7968 \\ \hline
			6 & 8.6875 & 7.6718 & 17.0781 & 15.1718 & 25.7656 & 23.8593 & 24.75 & 22.8437 & 24.1875 \\ \hline
			7 & 6.0937 & 5.4062 & 17.0781 & 15.1718 & 23.1718 & 21.2656 & 22.4843 & 20.5781 & 22.6093 \\ \hline
			8 & 4.4062 & 4.0625 & 17.0781 & 15.1718 & 21.4843 & 19.5781 & 21.1406 & 19.2343 & 21.25 \\ \hline
			9 & 3.6875 & 3.5 & 17.0781 & 15.1718 &	20.7656 & 18.8593 &	20.5781 & 18.6718 & 20.5937 \\ \hline	 
		\end{tabular}
	\end{adjustbox}
	\caption{Validation of Cor. \ref{cor_superposition} with two inputs $w_1$ as reference at frequency of $40\ Hz$ and $w_2$ as disturbance at frequency of $40\ Hz$, both independently given as input and also combined. $||e||_\infty$ is provided in units of $0.1 \mu m$ and is also measured for the base-linear system.}
	\label{tab_error_superposition}
\end{table*}

Finally, Cor. \ref{cor_superposition} related to the use of superposition with the concept of the virtual harmonic separator is verified in practice with the use of two exogenous inputs. According to Cor. \ref{cor_superposition}, if the error seen independently with one of the inputs (say $w_1$) is quite small compared to the error seen independently with the other (say $w_2$), then the first input $w_1$ is handled by the base-linear system. Several trials are conducted with $w_1$ as reference and $w_2$ as disturbance for difference amplitudes. Within each trial, the error is obtained for independent application of $w_1$ and $w_2$ and tabulated in the second and fourth columns of Table. \ref{tab_error_superposition} respectively. Additionally, the error is also obtained for the base-linear system (by setting $\gamma = 1$) for both inputs independently and tabulated in the third and fifth columns respectively. And finally, both $w_1$ and $w_2$ are simultaneously added to obtain the overall error as tabulated in the last column. An analysis of these numbers indicates that for trials 1,2 and 3, the measured $||e||_\infty$ follows Cor. \ref{cor_superposition} with the values closely matching the seventh column, where the second input $w_2$ is handled by the base-linear system. Similarly, with trails 7, 8 and 9, $w_1$ is handled by the base-linear system, with the values closely matching the eighth column. For trails 4 and 6, as the error by each source becomes comparable, the system moves away from Cor. \ref{cor_superposition} and this is even more clearly seen with trial 5. From these preliminary experiments, it appears that Cor. \ref{cor_superposition} holds reasonably well for peak error by one signal being up-to half the peak error of an additional signal. However, more experiments are required for verification. Additionally, it must be noted that the use of the same frequency for $w_1$ and $w_2$, albeit one added as reference and one as disturbance, along with the fact that the peak error of each signal matched in phase meant that the peak errors could be directly added and verified. Else, the phase of the individual error harmonics must be considered and added to obtain an estimate.

\section{Analysis for loop-shaping}
\label{section_Analysis}

From the elaborate simulation results as well as additional results from practice, it is clear that (i) existing DF based prediction is inaccurate to the extent that it cannot be used to optimally tune these controllers for performance (ii) proposed novel HOSIDF based method while not completely accurate, is still capable of predicting difference in performance and is more suited for analysis. In this section, we provide some remarks regarding the accuracy of the prediction method as well as some general tuning guidelines.

\begin{figure}
	\centering
	\includegraphics[trim = {0.0cm 0cm 0.0cm 0cm}, width=1\linewidth]{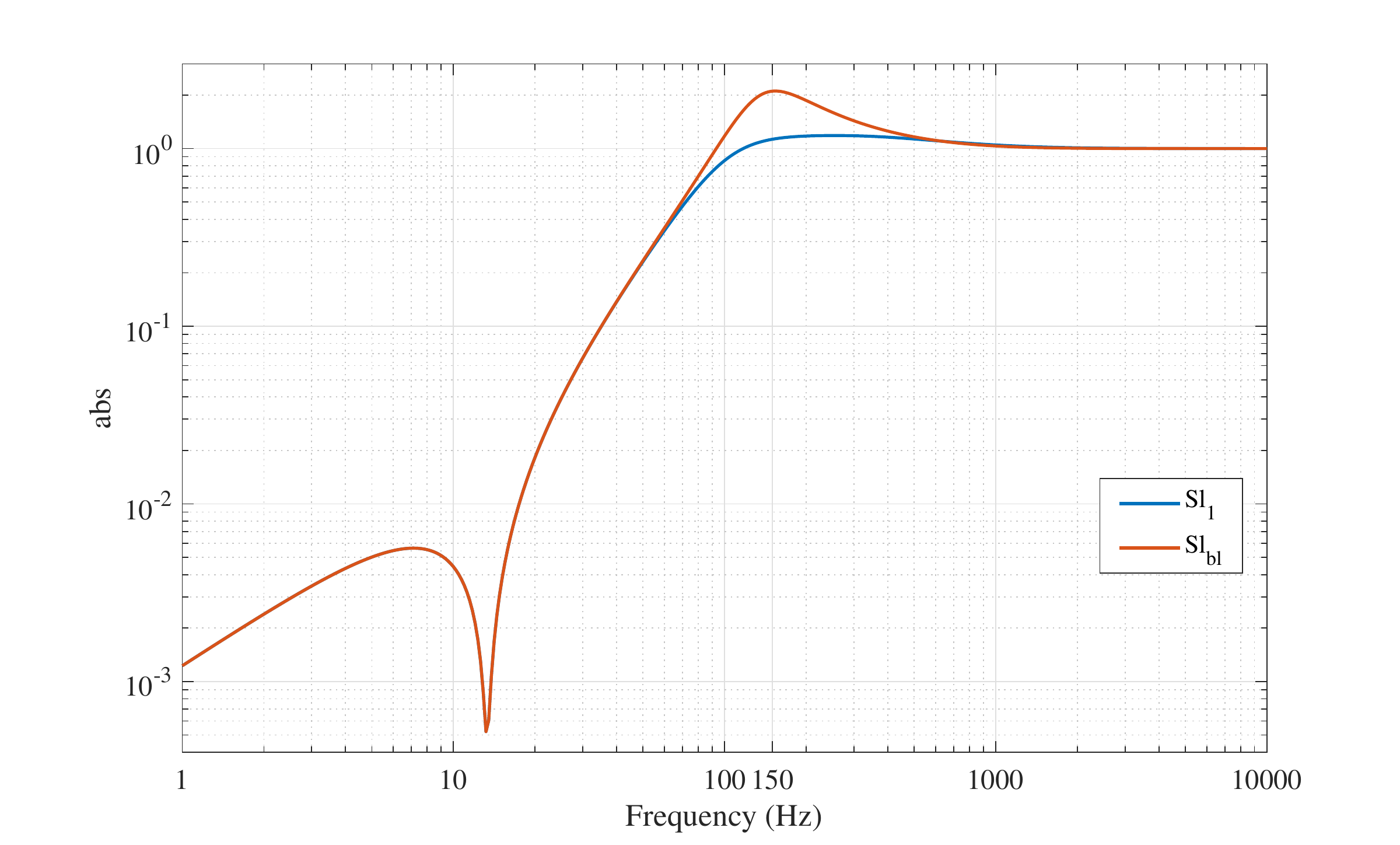}
	\caption{Sensitivity plots $Sl_1$ as defined by Eqn. \ref{eq_senseol_n} and $Sl_{bl}$ as defined by Eqn. \ref{eq_senseol_bl} for $\mathcal{C}_{02}$ based $\mathcal{RCS}$.}
	\label{fig_sensitivity_peaks}    
\end{figure}

While the PER plots of Figs. \ref{fig_sensitivity_PER_CCgLp_C2_C6} and \ref{fig_process_sensitivity_PER_CCgLp_C2_C6} show a very close match in the prediction error of the different controllers, this is not the case in Fig. \ref{fig_sensitivity_process_sensitivity_PER_comp_CCgLp}, where the prediction accuracy is vastly different. This trend is not only true for the PER plots of HOSIDF based prediction, but also DF based one. These can both be explained by an analysis of the Theorems. \ref{th_ol_to_cl_r}, \ref{th_ol_to_cl_d} and \ref{th_ol_to_cl_n}. In all the cases, while the $E_1(\omega)$ is dependent on $Sl_1(\omega)$, $E_{n\geq 2}(\omega)$ are dependent on $Sl_{bl}(\omega)$. While the first term is the sensitivity function purely based on DF, the second is based on the base-linear system. Since reset controllers are designed to increase the PM, the peak of sensitivity is higher for the second as seen in Fig. \ref{fig_sensitivity_peaks}. From Table. \ref{tab_CgLp_controllers}, we can see that while $\mathcal{C}_{10}$ and $\mathcal{C}_{01}$ have the same $\phi_{CgLp}$ resulting in comparable relative higher-harmonic magnitudes, the PM is different resulting in a huge difference in peak of $Sl_{bl}$. Since the larger peak results in larger magnitudes of $E_{n\geq 2}$, this results in a large deviation in the PER plots of DF based prediction. Additionally, the same large magnitudes of harmonics in error also influence the extent to which Assumptions 2 - 3 are violated resulting also in large prediction errors. This explains the large PER values for $\mathcal{C}_{01}$ compared to $\mathcal{C}_{10}$ in \ref{fig_sensitivity_process_sensitivity_PER_comp_CCgLp}.

From the perspective of tuning and performance, we seek to achieve the performance indicated by DF with appropriate suppression of the harmonics. Apart from $Sl_{bl}(n\omega)$, the harmonics are dependent on $L_n(\omega) = H_n(\omega)P(n\omega)$. It is trivial that a reduction of $H_n$ would result in better performance. Using this, we provide the guidelines below.
\begin{itemize}
	\item Given a stable base-linear system wherein a CgLp compensator has to be designed for optimal performance, choose the CgLp with the lowest $H_n$ at the required frequency.
	\item In general, since controllers are not designed for optimal performance at a single frequency, a weighted matrix can be used to calculate the CgLp configuration which is best matched. This can be considered for future work.
	\item The PM of the base-linear system determines the peak of $Sl_{bl}$ and hence a larger PM for the base-linear system results in better prediction and lower PER values. Hence, unlike retaining the given base-linear system as noted in the first point, if this can also be redesigned, i.e., linear part of $\mathcal{R}$ can be designed, then $Sl_{bl}$ has to also be considered and added to the optimisation cost function.
\end{itemize}

\section{Conclusions}
\label{section_Conclusions}
Reset controllers have shown great promise in overcoming the limitations of linear control and providing significant performance improvement. However, existing DF based loop-shaping and prediction cannot be used for precision control. Hence, we have provided (i) the extension of DF in the form of HOSIDF of reset controllers for accurate analysis in open-loop (ii) a novel prediction method based on the HOSIDFs with the introduction of the concept of a virtual harmonic separator for these systems. The prediction accuracy of the new method is seen to be significantly better and the capability to predict trends as seen with the practical results shows the potential of this method to be used for optimal tuning. Additionally, based on the results and the novel method developed, we have provided tuning guidelines for manual tuning of this family of controllers.

To further improve these prediction methods, in the next step, we strive to be able to estimate the PER values based on the magnitude of the harmonics and the extent to which Assumptions 2 and 3 are violated. Also, while we provide some basic guidelines for tuning in this work, it is also necessary to investigate new architectures capable of providing required DF in open-loop with suppressed HOSIDFs for improved performance. The presented methods while not perfect, provide a significant step forward for the design and analysis of these systems and moves towards ensuring greater utilization of these controllers in the high-tech industry setting.

\section*{Acknowledgement}
This work was supported by NWO, through OTP	TTW project \#16335.

\section*{}
Conflict of interest - none declared.
	
\bibliographystyle{unsrtnat}
\bibliography{ref}
	
\end{document}